\documentclass[fleqn,usenatbib]{mnras}

\usepackage[T1]{fontenc}
\DeclareRobustCommand{\VAN}[3]{#2}
\let\VANthebibliography\thebibliography
\def\thebibliography{\DeclareRobustCommand{\VAN}[3]{##3}\VANthebibliography}

\usepackage{graphicx}	
\usepackage{amsmath}	
\usepackage{tabularx}
\usepackage{amssymb}	
\usepackage{newtxtext,newtxmath}

\newcommand{\bs}{\boldsymbol}

\newcommand{\new}[1]{\textbf{#1}} 

\newcommand{\feh}{\ensuremath{\mathrm{[Fe/H]}}}
\newcommand{\mgfe}{\ensuremath{\mathrm{[Mg/Fe]}}}
\newcommand{\alfe}{\ensuremath{\mathrm{[Al/Fe]}}}

\newcommand{\Msun}{\ensuremath{\mathrm{M}_\odot}}
\newcommand{\kms}{\ensuremath{\mathrm{km\,s^{-1}}}}
\newcommand{\Eos}{Eos}

\title[Eos: a merger-induced starburst]{Eos: a merger-induced starburst at the dawn of Galactic low-$\alpha$ disc formation}

\author[H.Y. Zhāng et al.]{
Han-Yuan Zhāng,$^{1,2}$\thanks{hz420@cam.ac.uk}
Vasily Belokurov,$^{1}$
Sarah G. Kane,$^{1}$
and João A. S. Amarante$^{3,4}$
\\
$^{1}$ Institute of Astronomy, University of Cambridge, Madingley Road, Cambridge CB3 0HA, UK; \\
$^{2}$ Max-Planck-Institut für Astronomie, Königstuhl 17, 69117 Heidelberg, Germany \\
$^{3}$ Department of Astronomy, School of Physics and Astronomy, Shanghai Jiao Tong University, 800 Dongchuan Road, Shanghai 200240,\\
People’s Republic of China; \\
$^{4}$ State Key Laboratory of Dark Matter Physics, School of Physics and Astronomy, Shanghai Jiao Tong University, Shanghai 200240,\\
People’s Republic of China; \\
}

\date{Accepted XXX. Received YYY; in original form ZZZ}
\pubyear{2026}

\begin{document}
\label{firstpage}
\pagerange{\pageref{firstpage}--\pageref{lastpage}}
\maketitle

\begin{abstract}
The last significant merger of the Milky Way, the \textit{Gaia}-Sausage-Enceladus (GS/E), is thought to have reshaped the early Galaxy and to mark the transition from the high-$\alpha$ to the low-$\alpha$ disc. The stars formed in situ during this merger offer a direct record of that epoch but are hard to identify. Using APOGEE\,DR17 abundances, we study Eos, a population of in-situ low-$\alpha$ stars on halo-like, non-rotating, eccentric orbits. \new{The metal-poor part of Eos has a nitrogen dispersion as large as that of the pre-disc Aurora population and up to 1.7 times that of the low-$\alpha$ disc at the same metallicity. Heating alone cannot produce this: the Splash, a disc heated onto halo orbits by the same merger and selected in the same way, shows no excess over its parent disc. This part of Eos was therefore born in poorly mixed gas. Eos is also chemically intermediate between the accreted GS/E debris and the coeval in-situ Milky Way stars, and stars with more GS/E-like chemistry move on more extended orbits.} Its stars are older than the bulk of the low-$\alpha$ disc but younger than the high-$\alpha$ disc and the Splash. We compare these observations with a merger-induced starburst seen in an Auriga simulation. There, a GS/E-like merger funnels the stripped progenitor gas into a mixing lane of intermediate metallicity, where stars form with intermediate abundances, a large chemical dispersion and a chemodynamical gradient set by their birth site, reproducing the observed signatures of Eos. We therefore interpret Eos as the surviving relic of the GS/E merger-induced starburst, born largely from the partially mixed gas bridging the two galaxies and marking the onset of the Milky Way's low-$\alpha$ disc. Eos thus provides a direct observational view of the star formation and galaxy evolution driven by the Milky Way's last significant merger.
\end{abstract}



\section{Introduction}
\label{sec:intro}

Mergers between galaxies are an effective way to produce a starburst, provided there is gas available (accreted, native or both) to convert into stars. Across the spectrum of mass ratios, from major to minor mergers, interactions usually lead to enhanced star formation \citep[see e.g.][]{Barnes1991, Barnes1992, Mihos1994, Mihos1996, Springel2000, Cox2008, Dimatteo2008, Renaud2014}, as seen observationally in interacting pairs and (ultra-)luminous infrared galaxies \citep[e.g.][]{SandersMirabel1996, Ellison2008}. These bursts are, however, short-lived, typically lasting only a few hundred Myr \citep[][]{Dimatteo2008}. They are driven by the tidal field of the encounter, which torques the gas, removes its angular momentum and funnels it towards the galactic centre, but which also compresses gas over extended volumes and along the tidal bridges and tails that link the interacting pair \citep[][]{Mihos1996, Renaud2014}. The star formation triggered by a merger is therefore not confined to the relaxed disc: a substantial fraction can occur in perturbed, out-of-equilibrium gas, so that the resulting stars are born with unusual chemistry and on dynamically hot orbits. Because such stars form from gas that is only partially mixed between the two galaxies, over a short interval and in a turbulent environment, they carry a distinctive chemodynamical fingerprint of the merger that, in principle, survives to the present day.

The Milky Way offers a uniquely resolved view of these processes. The last significant merger of the Milky Way, the \textit{Gaia}-Sausage-Enceladus (GS/E) merger \citep{Belokurov2018, Helmi2018}, took place $\approx10$--$12$\,Gyr ago and deposited a radially anisotropic debris cloud that dominates the inner stellar halo \citep[][]{Fattahi2019, Naidu2020, Wu2022}.

Recent measurements suggest that the merger was significant but probably not major. A merger is conventionally called major when the mass ratio of the two galaxies exceeds 1:4, and minor when it lies between 1:4 and 1:10 \citep{Lotz2011, RodriguezGomez2015, RodriguezGomez2016}. The GS/E debris has a stellar mass of $\approx1.5$--$2.3\times10^8$\,M$_\odot$ \citep{Lane2023, Lane2026}, and selection-corrected Gaia star counts give $\approx2.4$--$4.9\times10^8$\,M$_\odot$ \citep{Kurbatov2024}. Even the highest of these values is well below the total stellar mass of the Galactic halo, $\approx1.0$--$1.4\times10^9$\,M$_\odot$ \citep{Deason2019}, which also includes debris from other dwarfs and in-situ stars heated out of the disc. The Milky Way had already formed $\sim10^{10}$\,M$_\odot$ of stars by then \citep{Snaith2014, Orkney2026b_}, so the stellar mass ratio was $\lesssim1{:}10$, below the threshold for a major merger. The total mass ratio is larger, because in dwarf galaxies stellar mass falls steeply with halo mass \citep{Hopkins2010}. The stellar-to-halo mass relation \citep{Moster2013} places GS/E in a halo of $\approx0.7$--$1.5\times10^{11}$\,M$_\odot$. The Milky Way halo was already massive by then. Its disc formed before GS/E arrived \citep{BK2022}, unusually early for a galaxy of its mass \citep{Semenov2024}, and simulated discs form only once the host halo reaches $M_{\rm 200c}\approx(6\pm3)\times10^{11}$\,M$_\odot$ \citep{Dillamore2024}. The mass of the old disc likewise implies a Milky Way halo of $\approx7\times10^{11}$\,M$_\odot$ already 12.5\,Gyr ago \citep{Orkney2026b_}. Against such a halo, the total mass ratio was $\approx1{:}4$--$1{:}9$. Two dynamical arguments independently place GS/E below 1:4. First, the pre-merger disc survived: $\alpha$-rich stars as old as 13.5\,Gyr still rotate, which in Auriga happens only after mergers below 1:4 \citep{Orkney2026b_}. Second, the GS/E debris stays at relatively high orbital energies \citep{Belokurov2023wrinkles}; in simulations, only light progenitors behave this way, while heavier ones sink deeper into the potential of the host \citep{Amarante2022, Orkney2026b_}.

Beyond this accreted debris, the GS/E merger left a marked imprint on the \emph{in-situ} Galaxy. It dynamically heated the pre-existing high-$\alpha$ disc into a slowly rotating, metal-rich halo component, the Splash \citep{Belokurov2020, DiMatteo2019, Dillamore2022}, which, together with the pressure-supported pre-disc population known as Aurora \citep{BK2022}, forms part of the local stellar halo. The recognition that the local stellar halo comprises both accreted debris and an in-situ, heated-disc component has a long history \citep[e.g.][]{NissenSchuster2010, Bonaca2017, Haywood2018, DiMatteo2019, Gallart2019}. The in-situ response to the GS/E merger provides a direct, star-by-star record of how a significant merger reshapes a galaxy like the Milky Way.

The clearest chemical fossil of this epoch is the $\alpha$-bimodality. In the \mgfe--\feh\ plane, the disc separates into a high-$\alpha$ (thick) and a low-$\alpha$ (thin) sequence \citep[e.g.][]{Bensby2014, Haywood2013, Nidever2014, Hayden2015}, and the transition between them has increasingly been associated with the GS/E merger. Many mechanisms have been proposed to produce the bimodality: the accretion of metal-poor gas that dilutes the interstellar medium and resets the chemical sequence \citep[e.g.][]{Chiappini1997}, a star-formation hiatus or quenching that allows SNe\,Ia to lower \mgfe\ at fixed metallicity, radially selective gas sinking driven by a retrograde merger \citep[][]{Khoperskov2021, Johnson2025}, and radial migration \citep[][]{SchoenrichBinney2009, Sharma2021, Zhang2025}. Hydrodynamical simulations increasingly tie the onset of the low-$\alpha$ sequence to a gas-rich merger \citep[][]{Grand2020, Ciuca2024, Agertz2021_VI, Renaud2021_VIII, Buck2020, Amarante2026, Genina2026}, although the bimodality can also emerge without such a merger, both in cosmological models \citep{Orkney2026a_bimodality} and through massive star-forming clumps in the early disc \citep{Clarke2019, BeraldoeSilva2020, Amarante2026}. How exactly the GS/E merger seeded the low-$\alpha$ disc, and whether the earliest low-$\alpha$ stars can be identified and used to date the event, remain open questions.

If the GS/E merger drove a starburst, it should have left behind an in-situ population formed from the merger-processed gas and bearing the fingerprints set out above: a chemistry intermediate between the metal-poor, low-$\alpha$ accreted dwarf and the metal-rich, high-$\alpha$ host; an enhanced and stochastic chemical dispersion inherited from the poorly mixed gas; and heated, halo-like orbits. \new{Of these, the chemical dispersion is the most direct test: intermediate abundances can arise from many enrichment histories, and halo orbits from the heating of a disc, but a wide abundance spread at fixed metallicity requires stars to form from poorly mixed gas.} Cosmological simulations of GS/E-like mergers predict exactly such populations. A burst of stars on halo orbits forms at the last pericentric passage \citep[][]{Grand2020, Ciuca2024, Orkney2026b_}, some of them along the gas bridge that links the two galaxies, and tidal compression can build a misaligned, metal-poor outer disc \citep[][]{Agertz2021_VI, Renaud2021_VIII}. Identifying this population in the Galaxy would turn the in-situ response to the GS/E merger into an observable, allowing the star formation and gas dynamics of the Milky Way's last significant merger to be studied directly.

A chemodynamically peculiar population in the local stellar halo has recently emerged as a candidate for the in-situ product of the merger-induced starburst. \citet{Myeong2022} applied a Gaussian mixture model to the chemistry and dynamics of highly eccentric local stars and isolated \Eos, a structure at the most metal-poor end of the low-$\alpha$ sequence with light-element abundances intermediate between those of the Milky Way disc and the GS/E. \citet{Davies2025} confirmed it as a distinct structure using non-negative matrix factorisation and noted that it splits into two branches in the \mgfe--\feh\ plane. \citet{An2025} identified a bimodal population of low-$\alpha$, high-[Na/Fe] stars that is plausibly the same structure, and interpreted it as the product of clumpy star formation triggered by the merger. However, settling the origin of Eos requires a quantitative comparison between its chemodynamical properties and those of a merger-induced starburst in simulations; this work fills that gap.

In this work, we characterise the chemodynamics of \Eos\ using APOGEE\,DR17 abundances and \textit{Gaia} astrometry, and we interpret it by comparison with the star formation triggered by a GS/E-like merger in the Auriga cosmological zoom-in simulation Au18. The paper is organised as follows. Section~\ref{sec:data} describes the data and the population selection. Section~\ref{sec:eos} presents the kinematic, chemical and age properties of \Eos. Section~\ref{sec:sims} examines the merger-induced starburst in Au18 and the stars formed along the resulting gas lane. Section~\ref{sec:comparison} compares our results with other works in the literature and discusses the implications for the formation of the low-$\alpha$ disc, and Section~\ref{sec:conclusions} summarises our conclusions.

\section{Data}
\label{sec:data}

\subsection{Data selection}
\label{sec:data_obs}

\begin{figure*}
\centering
\includegraphics[width=\textwidth]{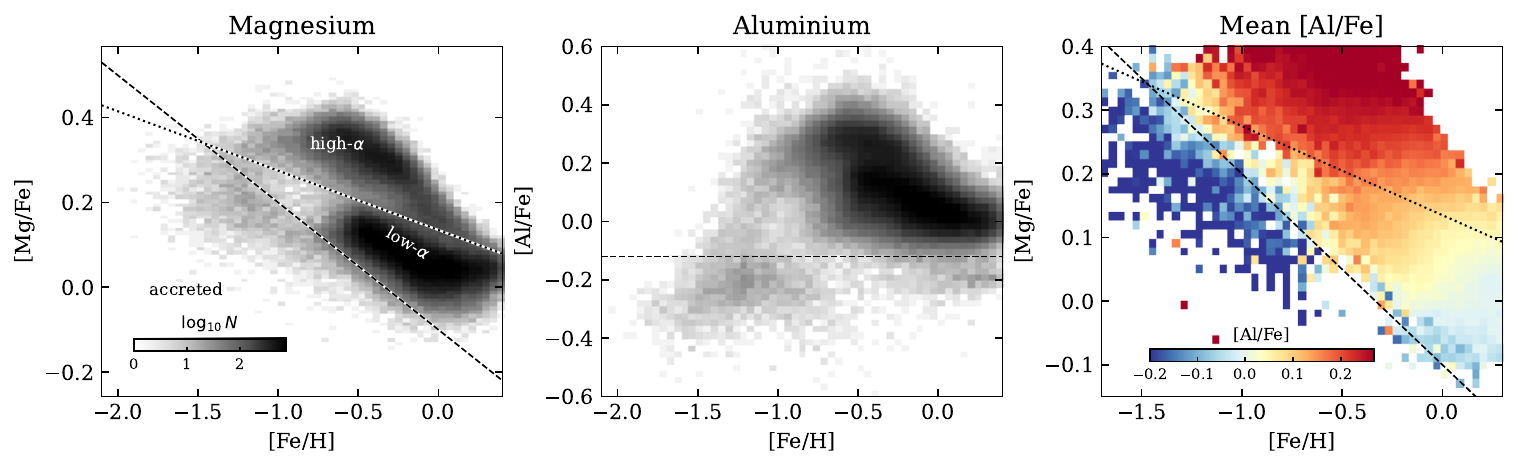}
\caption{Chemistry of the base APOGEE\,DR17 sample and the population selection.
\textit{Left:} Logarithmic stellar density in the \mgfe--\feh\ plane. The dashed line ($\mgfe=-0.30\,\feh-0.10$) separates the accreted debris from the in-situ populations, and the dotted line ($\mgfe=-0.14\,\feh+0.135$) separates the high-$\alpha$ and low-$\alpha$ sequences. \textit{Middle:} Logarithmic stellar density in the \alfe--\feh\ plane; the dashed line marks the in-situ cut $\alfe=-0.12$. \textit{Right:} The \mgfe--\feh\ plane colour-coded by the mean \alfe\ of stars with $-0.75\times10^5 < E < -0.4\times10^5$~(km\,s$^{-1}$)$^2$ (the orbital-energy band of the local halo in which Eos dominates; $72$ per cent of Eos stars fall in this window, against $28$ per cent of the low-$\alpha$ disc). The accreted/in-situ boundary coincides with the transition from sub-solar to enhanced aluminium, reflecting the delayed Al enrichment of the accreted dwarf.}
\label{fig:mg_al}
\end{figure*}

We use data from the 17th data release (DR17) of APOGEE \citep[][]{Majewski2017, Abdurrouf2022}. The chemical abundances are
the ASPCAP \citep{GarciaPerez2016} measurements from the \texttt{allStarLite} catalogue. 
The Galactocentric coordinates, velocities, integrals of motion and orbital properties of the stars, as well as their uncertainties, are taken from the astroNN value-added catalogue \citep[see][]{Leung2019,Mackereth2018}, which assumes the \texttt{MWPotential2014} gravitational potential of \citet{Bovy2015}. Both catalogues make use of the {\it Gaia} Early Data Release 3 photometry and astrometry \citep[][]{EDR3_summary, EDR3_validation, EDR3_photometry, EDR3_astrometry}. We follow the selection procedure of \citet{BK2022} to build a clean sample of Milky Way red giants. In brief, we remove duplicates from the APOGEE DR17 catalogue, discard stars in and around known Galactic satellites and
globular clusters, and retain only giants ($\log g < 3$) with high-quality
kinematics (velocity uncertainties $<50\,\kms$) and chemistry (abundance
uncertainties $<0.2$\,dex). We also remove stars with problematic spectra using the following flags: \texttt{STAR\_BAD, TEFF\_BAD, LOGG\_BAD, VERY\_BRIGHT\_NEIGHBOR, LOW\_SNR, PERSIST\_HIGH, PERSIST\_JUMP\_POS, PERSIST\_JUMP\_NEG, SUSPECT\_RV\_COMBINATION}. As a precaution against distance systematics, we further limit the sample to heliocentric distances below $15$\,kpc. This yields a base sample of $\sim187\,000$ stars.

Figure~\ref{fig:mg_al} summarises the [Fe/H], [Mg/Fe] and [Al/Fe] properties of the selected sample and defines the populations used throughout the paper. We use simple cuts to carve out three distinct populations: i) the accreted stars, predominantly GS/E debris \citep[][]{Belokurov2018, Helmi2018}, which show the fastest decline in [Mg/Fe], starting at [Fe/H]$\approx-1.3$; ii) the high-$\alpha$ population, corresponding to the Milky Way's early disc as well as the Galaxy's pre-disc phase, known as {\it Aurora} \citep[see][]{BK2022}; and iii) the low-$\alpha$ population, corresponding to the late stage of disc formation. The left panel shows the [Fe/H]--[Mg/Fe] distribution of the sample, with the lines separating the accreted, high-$\alpha$ and low-$\alpha$ populations. At high metallicity ([Fe/H]$>-0.5$), this third group has the lowest [Mg/Fe] ratios ([Mg/Fe]$<0.1$), but at lower metallicity ([Fe/H]$<-1$) its $\alpha$ abundances are intermediate between those of the accreted and the high-$\alpha$ stars. The middle panel of Figure~\ref{fig:mg_al} shows the stellar density in the plane of [Al/Fe] and [Fe/H]. Three distinct populations are readily identified here: the accreted halo, dominated by the GS/E debris, with low [Al/Fe] values not exceeding [Al/Fe]$\approx-0.1$; and the high- and low-$\alpha$ discs, with [Al/Fe] values typically above $-0.1$, which also include the in-situ stars of Aurora \citep[see][]{BK2022} at [Fe/H]$<-1$. The right panel shows the [Fe/H]--[Mg/Fe] plane coloured by the mean [Al/Fe] of stars with orbital energy $-0.75\times10^5 < E < -0.4\times10^5$~(km\,s$^{-1}$)$^2$. The accreted/in-situ boundary coincides with a switch from low (blue) to high (red) aluminium, showing that [Al/Fe] separates these two stellar populations.

\subsection{Preview of the disc and halo chemodynamics}

\begin{figure*}
    \centering
    \includegraphics[width=\linewidth]{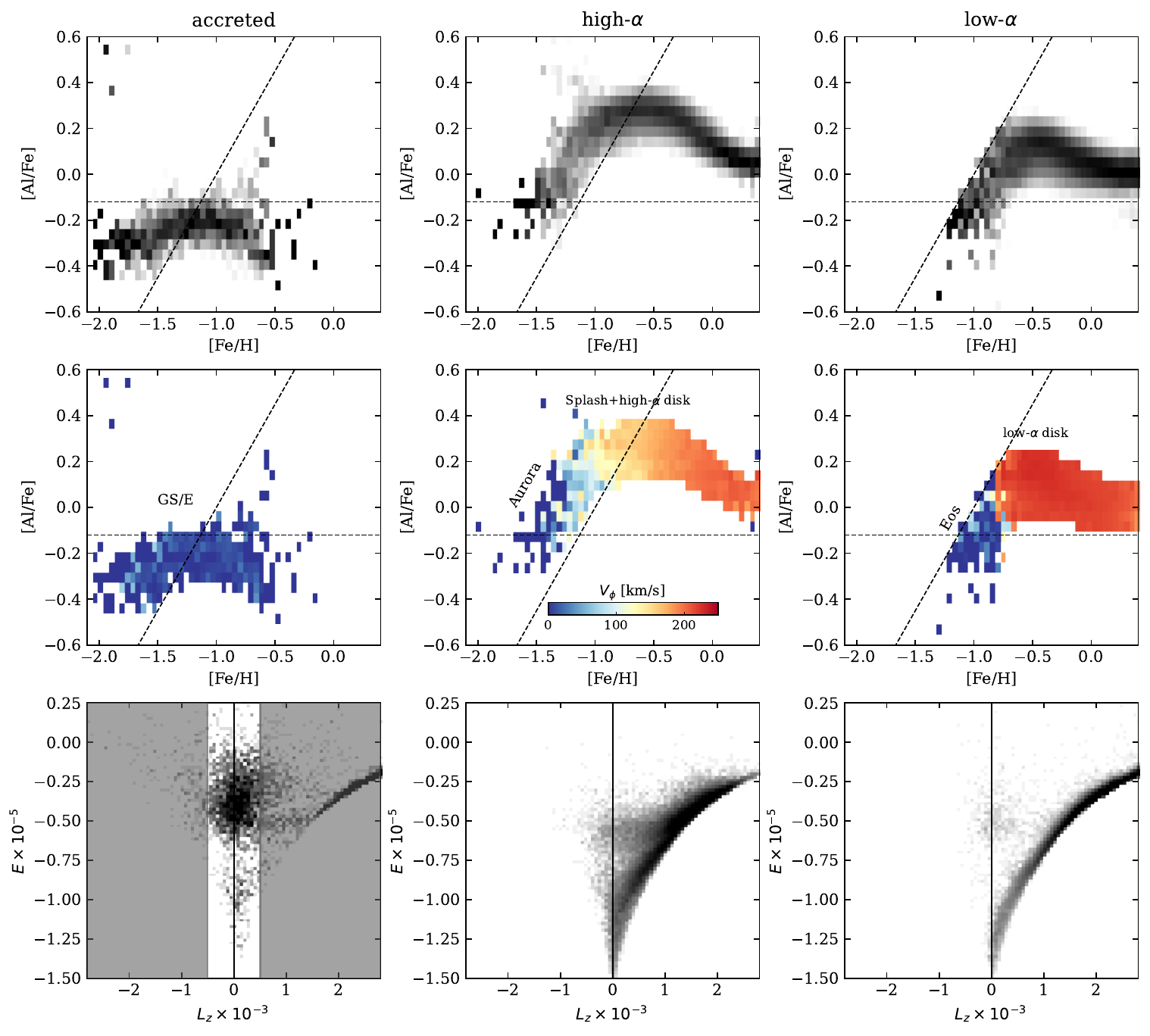}
    \caption{Chemodynamical properties of the high-$\alpha$, low-$\alpha$ and accreted populations. Columns show the accreted, high-$\alpha$ and low-$\alpha$ samples (left to right). \textit{Top:} Column-normalised density in the [Fe/H]--[Al/Fe] plane. The horizontal dashed line marks the in-situ cut [Al/Fe]$=-0.12$, and the diagonal dashed line the additional cut [Al/Fe]$=0.9$[Fe/H]$+0.9$ applied to the low-$\alpha$ sample to remove high-$\alpha$ contaminants. In the top and middle accreted panels, stars are restricted to $|L_z|<500$~kpc\,km\,s$^{-1}$; the bottom-left panel instead shows the full chemically selected accreted sample, with the excluded $|L_z|>500$ region shaded. \textit{Middle:} The same [Fe/H]--[Al/Fe] plane coloured by the mean azimuthal velocity of the stars, $v_\phi$. The colour change in the high-$\alpha$ sample marks the transition from Aurora to the Splash and the high-$\alpha$ thick disc. \textit{Bottom:} The integrals-of-motion ($E$--$L_z$) plane for the three populations. In the bottom right panel, the small cloud at $L_z\approx0$ and $E\approx-0.5\times10^5$~(km\,s$^{-1}$)$^2$ is the Eos population, which is dynamically distinct from the rotating low-$\alpha$ disc. }
    \label{fig:3pops}
\end{figure*}

Figure~\ref{fig:3pops} highlights the differences in the behaviour of [Al/Fe] as a function of metallicity for the three stellar populations selected in the left panel of Figure~\ref{fig:mg_al}. The top panels of Figure~\ref{fig:3pops} give the column-normalised density for each sample of stars. In addition, we restrict the accreted population (left) to $|L_z|<500$ kpc~km~s$^{-1}$. This angular-momentum mask is shown in the bottom-left panel of the figure, where the background density shows the $E$--$L_z$ distribution of all stars passing only the chemical selection. As the top panels of Figure~\ref{fig:3pops} demonstrate, the accreted, high-$\alpha$ and low-$\alpha$ stars follow distinct, well-defined and relatively narrow tracks in the [Al/Fe] vs [Fe/H] plane. As discussed in \citet{BK2022}, Al yields are strongly metallicity-dependent \citep[see e.g.][]{Kobayashi2006}, which delays Al production relative to that of the $\alpha$-elements. Because dwarf galaxies build up their stellar mass more slowly than the Milky Way, they reach the metallicity needed for efficient Al production too late, i.e. around the time SNe\,Ia start to contribute iron to the ISM, which suppresses the increase in the [Al/Fe] ratio. This can be seen in the top-left panel of the figure, which shows the accreted population. The accreted [Al/Fe] ratios rise very slowly and do not exceed [Al/Fe]${\approx}-0.1$. This nearly flat pattern contrasts with the [Al/Fe] sequence of the high-$\alpha$ population (top centre), which rises quickly above [Al/Fe]${\approx}-0.1$ and reaches [Al/Fe]${\approx}+0.3$ around [Fe/H]${\approx}-0.8$. 

The low-$\alpha$ sequence behaves differently from both the accreted and the high-$\alpha$ stars. First, its rise in [Al/Fe] is delayed: the sequence crosses the [Al/Fe]$=-0.1$ mark at [Fe/H]${\approx}-1$. Second, its [Al/Fe] increases with metallicity more slowly than that of the high-$\alpha$ stars. We use the [Al/Fe] trends of the three populations to build a clean sample of low-$\alpha$ stars. Below [Al/Fe]$=-0.12$, the accreted and in-situ populations cannot be cleanly separated; therefore, in the analysis below (unless specified otherwise), we focus on the in-situ stars with [Al/Fe]$>-0.12$ (horizontal dashed line in the top and middle rows). Note also that the high- and low-$\alpha$ sequences come close together at $-1.2<$[Fe/H]$<-0.7$. To minimise the contamination of the low-$\alpha$ sequence by high-$\alpha$ stars, we apply an additional cut, [Al/Fe]$<0.9[\rm Fe/H]+0.9$, shown as the diagonal dashed line in the top and middle rows. Typical low-$\alpha$ (sometimes referred to as thin-disc) stellar samples are limited to [Fe/H]$>-0.6$. Here, we extend this population by some 0.4 dex, to [Fe/H]${\approx}-1$. 

The middle and bottom panels of Figure~\ref{fig:3pops} show the orbital properties of these three populations, which are selected mainly by their chemistry. The accreted stars (left column) show little rotation and are restricted to $|L_z|<500~\rm kpc~km~s^{-1}$, but they reach the highest total energies and (in this sample) mostly have $E>-0.6\times10^5$~(km\,s$^{-1}$)$^2$. The high-$\alpha$ sample (centre column) contains both the early (thick) disc of the Galaxy and the in-situ halo, which spans a wide range of angular momentum. Note how the high-$\alpha$ stars stretch from close-to-zero to high positive $v_\phi$ with increasing metallicity, marking the spin-up epoch of the Milky Way at $-1.3\lesssim[\rm Fe/H]\lesssim-1$. The high-$\alpha$ stars on halo-like orbits belong to Aurora \citep[][]{BK2022} and the Splash \citep[]{Belokurov2020}. Finally, the middle and bottom right panels of Figure~\ref{fig:3pops} give the orbital properties of the low-$\alpha$ sample. As expected, most of the low-$\alpha$ stars lie in the fast-rotating (thin) disc of the Milky Way. Note, however, the small but noticeable group of stars with zero angular momentum at $E\approx-0.5\times10^5$~(km\,s$^{-1}$)$^2$. These low-$\alpha$ stars on halo-like orbits are distinct both from the accreted halo, because they occupy a different energy range, and from the in-situ halo, because they have a much narrower $L_z$ distribution, completely disconnected from that of the disc. The middle-right panel of Figure~\ref{fig:3pops} shows the same separation along the chemical track. Unlike the gradual spin-up of the high-$\alpha$ stars, $v_\phi$ in the low-$\alpha$ sample jumps from close to zero to high values: below [Fe/H]$\approx-0.85$ almost all low-$\alpha$ stars with [Al/Fe]$>-0.12$ are non-rotating, and the rotating disc takes over within $\approx0.1$\,dex. \citet{Viswanathan2025} also find a sharp transition in their low-$\alpha$ sample at [M/H]$\approx-1$, which they attribute to GS/E debris dominating its metal-poor end. Here the accreted stars are removed by the [Al/Fe] cut, and the non-rotating stars that remain are in situ. This population was dubbed \emph{Eos} by \citet{Myeong2022} and re-analysed by \citet{Davies2025}. 

\section{Low-$\alpha$ disc stars with halo-like orbits}
\label{sec:eos}

Above, we described a simple set of selection cuts that yields a highly pure sample of stars dominated by the Galactic low-$\alpha$ (or thin) disc. A small fraction of the stars in this group appear to have zero mean angular momentum, and thus show halo-like orbits despite having chemistry similar to that of the low-$\alpha$ disc, as previously reported \citep{OrtigozaUrdaneta2023, Filion2025, Borbolato2026}. Here, we investigate this peculiar Galactic population further.

\subsection{Kinematics and dynamics}
\label{sec:eos_kinematics}

\begin{figure*}
    \centering
    \includegraphics[width=\linewidth]{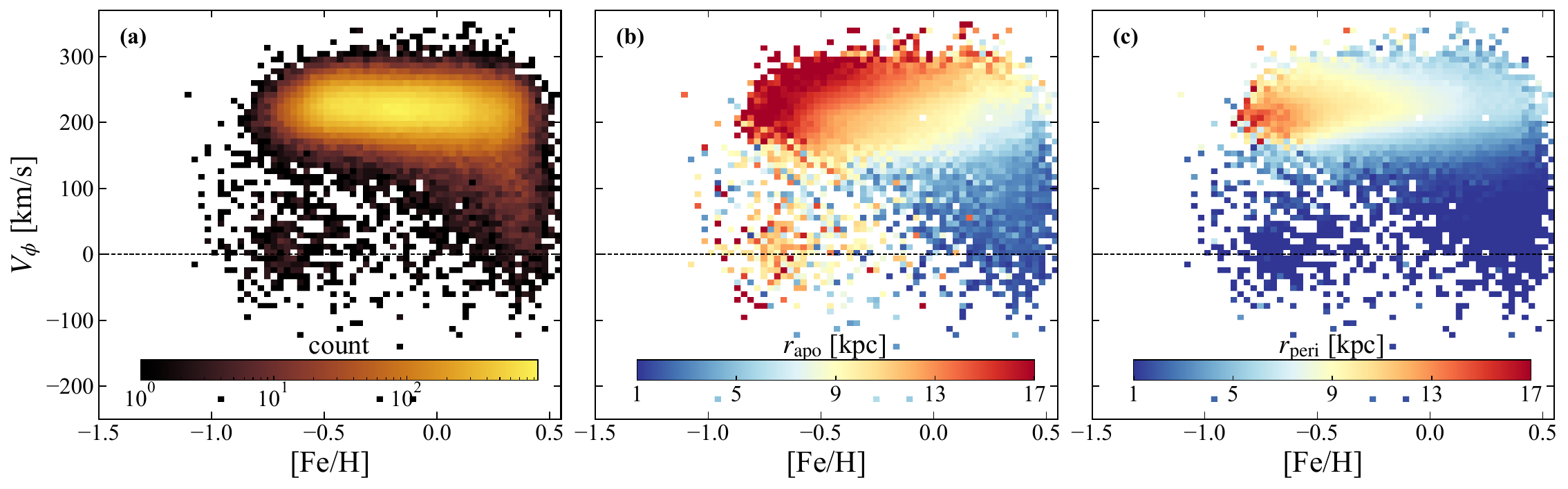}
    \caption{Kinematics of the in-situ low-$\alpha$ sample in the [Fe/H]--$v_\phi$ plane. \textit{Left:} Stellar number density per pixel. \textit{Middle} and \textit{Right:} The same plane coloured by the mean apocentric radius $r_{\rm apo}$ and pericentric radius $r_{\rm peri}$, respectively. In all panels, the dashed line marks $v_\phi=0$. The dominant, fast-rotating thin disc at [Fe/H]$>-0.5$ shows a smooth gradient of $r_{\rm apo}$ and $r_{\rm peri}$ with metallicity, reflecting the radial metallicity gradient of the disc. Eos occupies the metal-poor, low-$v_\phi$ corner ([Fe/H]$\lesssim-0.5$, $|v_\phi|\lesssim80~\kms$) on highly eccentric orbits with large $r_{\rm apo}$ and small $r_{\rm peri}$, and is separated from the rotating disc by a clear gap in $v_\phi$ at fixed metallicity.}
    \label{fig:vphi}
\end{figure*}

\begin{figure*}
    \centering
    \includegraphics[width=\linewidth, height=0.34\linewidth]{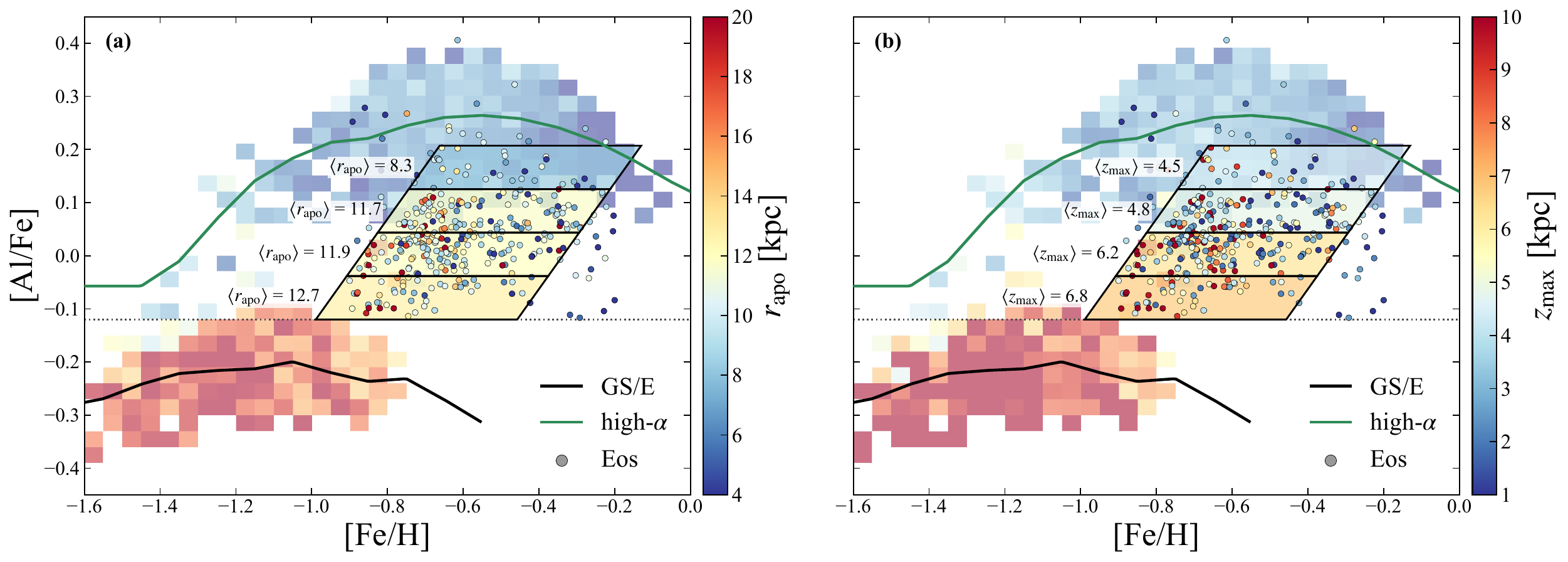}
    \caption{Coupling between the chemistry and orbits of Eos in the [Fe/H]--[Al/Fe] plane, coloured by the mean apocentric radius $r_{\rm apo}$ \textit{(left)} and the maximum vertical height $z_{\rm max}$ \textit{(right)}. The background image shows all stars on halo orbits ($e>0.7$ or $L_z<0$; $n=8083$); the overplotted points show the $353$ Eos stars, and the filled bands their mean $r_{\rm apo}$ and $z_{\rm max}$ in bins of [Al/Fe], as annotated. The four [Al/Fe] cells are of equal width from the in-situ cut $-0.12$ to the $95$th percentile of the Eos [Al/Fe] distribution; their slanted edges are lines of unit slope at the $5$th and $95$th percentiles of [Fe/H]$-$[Al/Fe], and the cell means use only the $308$ Eos stars that fall inside the cells. The black and green curves trace the median tracks of the accreted GS/E and the high-$\alpha$ populations, respectively, and the dotted line marks the in-situ cut [Al/Fe]$=-0.12$. Eos is intermediate between the GS/E debris (large $r_{\rm apo}$ and $z_{\rm max}$) and the in-situ halo, and its orbital extent decreases as its chemistry becomes more Milky Way-like (higher [Al/Fe]).}
    \label{fig:al_fe_orb}
\end{figure*}

\begin{figure*}
    \centering
    \includegraphics[width=\linewidth]{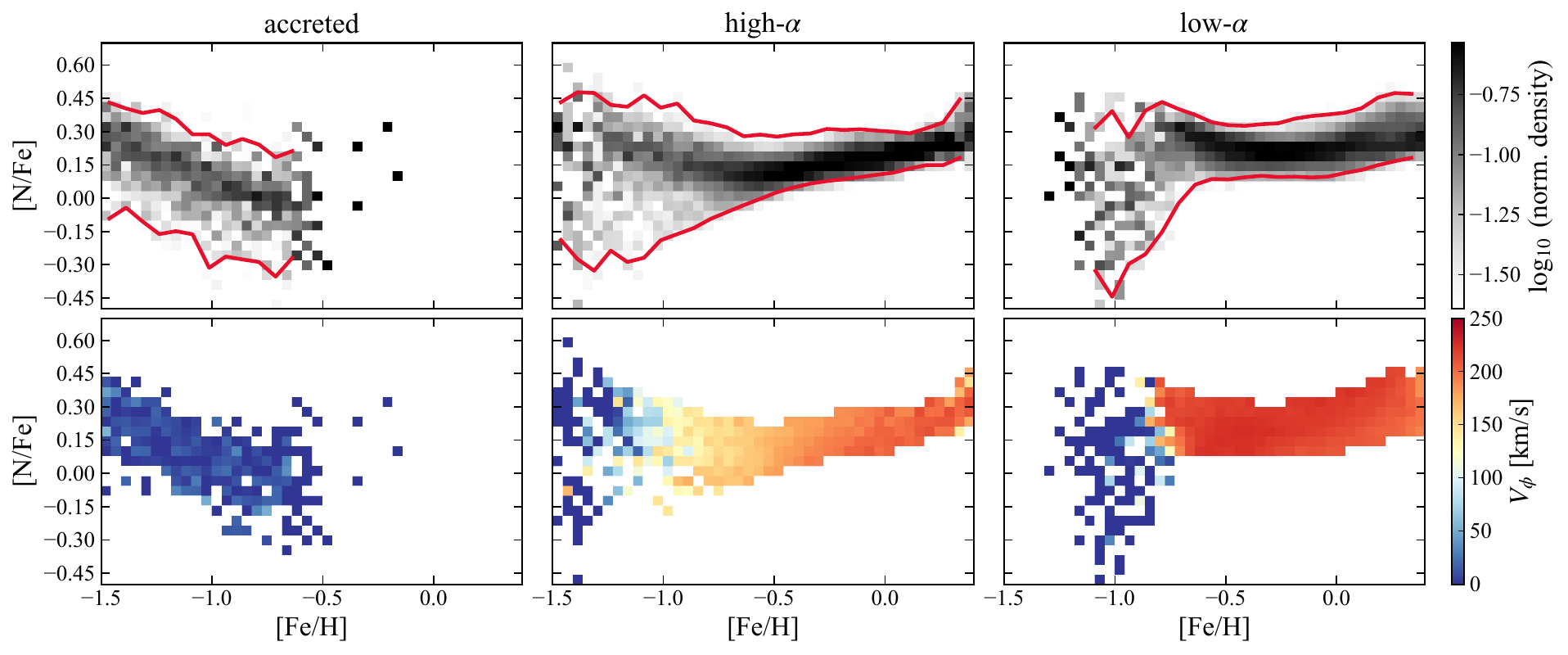}
    \caption{Nitrogen abundances of the accreted, high-$\alpha$ and low-$\alpha$ populations (left to right columns). \textit{Top:} Column-normalised density in the [Fe/H]--[N/Fe] plane; the red curves mark the 5th and 95th percentiles of [N/Fe] at fixed metallicity. \textit{Bottom:} The same planes coloured by the median $v_\phi$; the low-$v_\phi$ stars correspond to Aurora in the high-$\alpha$ column and to Eos in the low-$\alpha$ column. Eos shows a markedly broader nitrogen distribution than the low-$\alpha$ disc and the GS/E debris, comparable to that of the pre-disc Aurora population.}
    \label{fig:N_Fe_obs}
\end{figure*}

Figure~\ref{fig:vphi} investigates the correlations between [Fe/H] and the azimuthal velocity, $v_\phi$, in the low-$\alpha$ sequence as a function of apocentric and pericentric distance. The left panel shows the density distribution of the low-$\alpha$ sample in the [Fe/H]--$v_\phi$ plane. The main population, at [Fe/H]$>-0.7$ and $v_\phi>150$\,\kms, is the low-$\alpha$ thin disc. The middle and right panels show the same plane coloured by the mean apocentric and pericentric radius. In the main low-$\alpha$ thin disc population, apocentric and pericentric radii correlate clearly with metallicity, reflecting the metallicity gradient of the disc. Pericentres of the rotating low-$\alpha$ population are typically between 4 and 15 kpc. At high metallicity, i.e. [Fe/H]$>0$, we note a considerable population with low and negative values of $v_\phi$. These stars, mostly of super-solar metallicity, have pericentres of order 1 kpc and apocentric distances (in an unbarred potential) of around 5 kpc. In other words, they belong to the bulge/bar region of the Galaxy, which has non-circular motions and lower rotation speeds than the disc.

Figure~\ref{fig:vphi} shows that Eos is clearly a chemokinematically distinct population from the rest of the low-$\alpha$ sample: it sits in the metal-poor, low-$v_\phi$ corner, with the majority of its stars at $[\rm Fe/H]\lesssim-0.5$ and $-80\lesssim v_\phi\lesssim80$\,\kms. Its stars are characterised by typically large apocentres ($\approx10~\rm kpc$) and small pericentres ($\approx1~\rm kpc$), i.e. highly eccentric orbits with $e\gtrsim0.8$. Eos therefore has the highest eccentricities in the low-$\alpha$ sample. In Figure~\ref{fig:vphi}, not only is there a paucity of stars at $v_\phi\approx0$ in the metallicity range between Eos and the bulge/bar, but the Eos population also appears well separated from the rotating low-$\alpha$ disc. At the same metallicity, $-1<$[Fe/H]$<-0.5$, there is a clear gap in the $v_\phi$ distribution between Eos and the low-$\alpha$ disc.

We further explore the link between the chemistry and dynamics of the Eos population in Figure~\ref{fig:al_fe_orb}. We select stars on halo-like orbits with $e>0.7$ or $L_z<0$. Among these, we select as Eos stars those that sit in the low-$\alpha$ regime in Figure~\ref{fig:mg_al} and have $-1<[\rm Fe/H]<-0.2$. The Eos stars are shown as points coloured by their $r_{\rm apo}$ or $z_{\rm max}$. The background shows the other halo stars in the Milky Way; these are also coloured by the mean $r_{\rm apo}$ or $z_{\rm max}$. The upper blue clumps, with relatively small $r_{\rm apo}$ and $z_{\rm max}$, are mainly composed of the in-situ Milky Way halo population, with Aurora on the metal-poor side ($[\rm Fe/H]\lesssim-1$) and the Splash on the metal-rich side. The green line shows the median [Al/Fe]--[Fe/H] evolution of the Milky Way's high-$\alpha$ population. In contrast, the lower red clump is mainly composed of accreted GS/E stars, which have higher $r_{\rm apo}$ and $z_{\rm max}$ than the in-situ Milky Way population; the black line shows its median. As we have seen in the top row of Figure~\ref{fig:3pops}, the chemistry of the Eos population is intermediate between that of the accreted GS/E population and that of the in-situ Milky Way population. As Figure~\ref{fig:al_fe_orb} demonstrates, not only its chemistry but also its dynamical properties are intermediate. We highlight this by binning the Eos stars into cells of $[\rm Al/Fe]$ and calculating the mean $r_{\rm apo}$ and $z_{\rm max}$ in each cell. The Eos population has $8\lesssim r_{\rm apo}\lesssim13~\rm kpc$ and $4\lesssim z_{\rm max}\lesssim7~\rm kpc$, both intermediate between the respective values for the GS/E and the Milky Way's in-situ halo population. It also shows a clear gradient, with smaller $r_{\rm apo}$ and $z_{\rm max}$ at higher [Al/Fe], i.e. towards Milky Way-like chemistry. This could imply that the Eos population formed from a mixture of GS/E and Milky Way gas, for example in a gas lane containing gas stripped from the GS/E and gas flowing out of the Milky Way, where stars with chemistry more similar to the GS/E also have more similar dynamics, and likewise for the Milky Way. 

\subsection{Nitrogen dispersion: a poorly mixed birth environment}
\label{sec:eos_nitrogen}

\begin{figure*}
    \centering
    \includegraphics[width=\linewidth]{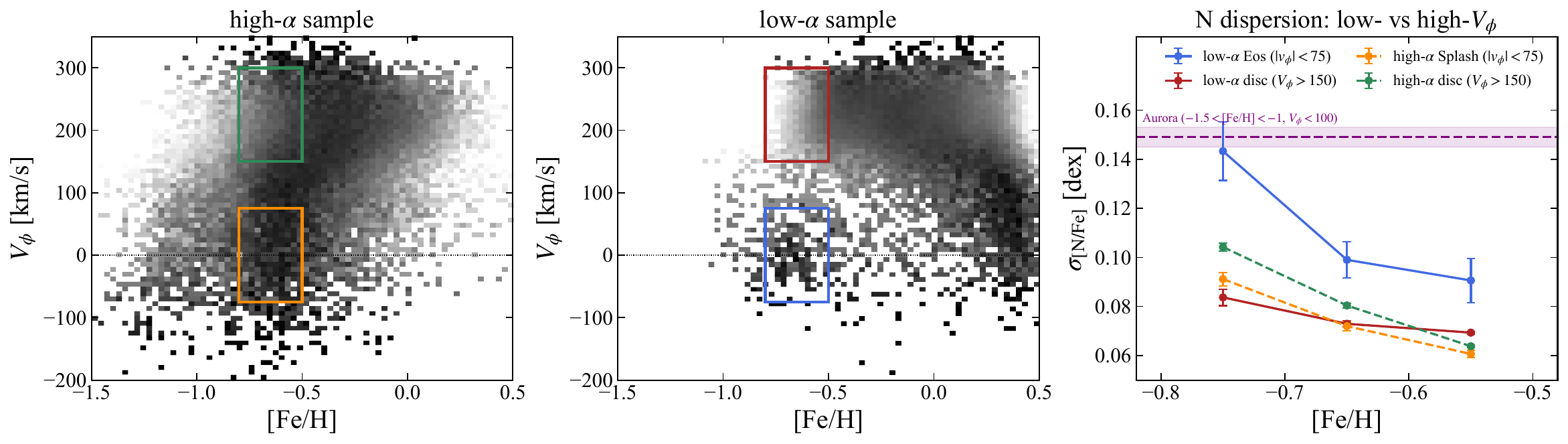}
    \caption{Nitrogen dispersion of Eos compared with the disc and the Splash. \textit{Left} and \textit{middle:} The row-normalised [Fe/H]--$v_\phi$ distributions of the high-$\alpha$ and low-$\alpha$ samples; the boxes mark the selections $-0.8<$[Fe/H]$<-0.5$ with $|v_\phi|<75~\kms$ and with $v_\phi>150~\kms$, coloured as the corresponding curves in the right panel. \textit{Right:} The error-deconvolved nitrogen dispersion $\sigma_{\rm [N/Fe]}$ as a function of metallicity for the four selections: low-$\alpha$ Eos (blue), low-$\alpha$ disc (red), high-$\alpha$ Splash (orange) and high-$\alpha$ disc (green); the purple band marks the Aurora benchmark ($-1.5<$[Fe/H]$<-1$, $v_\phi<100~\kms$). The Eos sample here is the low-$\alpha$ stars with $-0.8<$[Fe/H]$<-0.5$ and $|v_\phi|<75~\kms$ ($n=210$), which differs from the canonical Eos selection of Section~\ref{sec:data}; the comparison samples are the low-$\alpha$ disc ($v_\phi>150~\kms$, $n=9305$), the Splash ($|v_\phi|<75~\kms$, $n=2154$), the high-$\alpha$ disc ($v_\phi>150~\kms$, $n=10\,912$) and Aurora ($n=587$). The dispersions are error-deconvolved with $\sigma/\sqrt{2N}$ uncertainties; the Aurora value drawn ($0.149$) is the raw $1.4826\times$MAD. \new{At fixed metallicity, Eos has a nitrogen dispersion up to 1.7 times that of the low-$\alpha$ disc and reaches the Aurora value at its metal-poor end. The Splash, a heated disc, has a dispersion no higher than that of its parent high-$\alpha$ disc, so heating onto halo orbits does not broaden the nitrogen distribution.}}
    \label{fig:N_dispersion}
\end{figure*}

Nitrogen is a sensitive probe of the birth environment: the spread of [N/Fe] at fixed metallicity reflects how well mixed the star-forming gas was, so a wide spread points to formation from poorly mixed gas. Figure~\ref{fig:N_Fe_obs} and Figure~\ref{fig:N_dispersion} investigate the nitrogen abundances of the low-$\alpha$ sample. The upper panels of Figure~\ref{fig:N_Fe_obs} show the column-normalised [Fe/H]--[N/Fe] distributions of the accreted, high-$\alpha$ and low-$\alpha$ samples, with the red lines denoting the 5th and 95th percentiles of the [N/Fe] distribution at fixed metallicity. The lower panels show the same plane coloured by azimuthal velocity. As in Figure~\ref{fig:3pops}, we do not apply the $[\rm Al/Fe]>-0.12$ cut to the low- and high-$\alpha$ samples in Figure~\ref{fig:N_Fe_obs}. As in the middle panels of Figure~\ref{fig:3pops}, the lower panels show the transition from low to high $v_\phi$ with increasing metallicity, which is gradual in the high-$\alpha$ sample and abrupt in the low-$\alpha$ sample. The low-$v_\phi$ part (blue) of the high-$\alpha$ sample corresponds to the in-situ halo, Aurora, and that of the low-$\alpha$ sample corresponds to Eos. Since no [Al/Fe] cut is applied here, the metal-poor end of the low-$\alpha$ panel may also include some GS/E stars. 
The right panels of Figure~\ref{fig:N_Fe_obs} show that \new{Eos has a broad nitrogen distribution, significantly wider than that of the low-$\alpha$ disc and also wider than that of the accreted GS/E population shown in the left panel. A similarly broad distribution is seen in the Aurora population, as shown in the middle panels} \citep[see also][]{BK2022, Belokurov2023, Kane2025}. However, the uncertainties of the [N/Fe] measurements may also correlate with metallicity, because the absorption features become weaker at lower metallicity, leading to larger errors. 

To remove the potential metallicity dependence of the [N/Fe] errors, we measure the nitrogen dispersion of Eos and low-$\alpha$ disc stars in fixed metallicity bins in Figure~\ref{fig:N_dispersion}. The left two panels present the row-normalised $[{\rm Fe/H}]-v_\phi$ distributions of the high- and low-$\alpha$ samples. In both samples, we select stars with $-0.8<[\rm Fe/H]<-0.5$ and either $-75<v_\phi<75$\,\kms\ or $v_\phi>150$\,\kms. At these metallicities, for the high-$\alpha$ sample the high-$v_\phi$ stars correspond to the high-$\alpha$ thick disc and the low-$v_\phi$ stars to the splashed high-$\alpha$ disc, whereas for the low-$\alpha$ sample the high- and low-$v_\phi$ stars correspond to the low-$\alpha$ thin disc and Eos, respectively. The right panel of Figure~\ref{fig:N_dispersion} shows the nitrogen abundance dispersion $\sigma_{[\rm N/Fe]}$ as a function of metallicity for the selected stars. The displayed dispersion is error-deconvolved by subtracting in quadrature the median [N/Fe] error of the stars in each bin. \new{At fixed metallicity, the nitrogen dispersion of Eos is clearly higher than that of the low-$\alpha$ disc: $\sigma_{\rm [N/Fe]}\approx0.14$ against $\approx0.08$~dex at [Fe/H]$\approx-0.75$, with the ratio falling to $\approx1.3$ at [Fe/H]$\approx-0.55$. At its metal-poor end, Eos reaches the value of Aurora ($\approx0.15$~dex). The high-$\alpha$ stars provide a control for this measurement. The Splash is the high-$\alpha$ disc heated onto halo orbits by the GS/E merger (\citealt{DiMatteo2019,Belokurov2020}, but see \citealt{Amarante2020} for an in situ formation mechanism), and it is selected here with the same cuts as Eos. Its nitrogen dispersion is no higher than that of the high-$\alpha$ disc at any metallicity ($0.09$ against $0.10$~dex at [Fe/H]$\approx-0.75$). Heating a disc onto halo orbits therefore does not broaden its nitrogen distribution, and selecting stars on halo orbits does not by itself produce a large dispersion. The excess in Eos must have been set at birth: Eos stars formed from less well mixed gas than the low-$\alpha$ disc, in an environment resembling that of the pre-disc Milky Way. As we show in Section~\ref{sec:eos_bifurcation}, this excess is carried by the metal-poor branch of Eos.} 

\subsection{Age distributions}
\label{sec:eos_age}

\begin{figure*}
    \centering
    \includegraphics[width=\linewidth, height=0.34\linewidth]{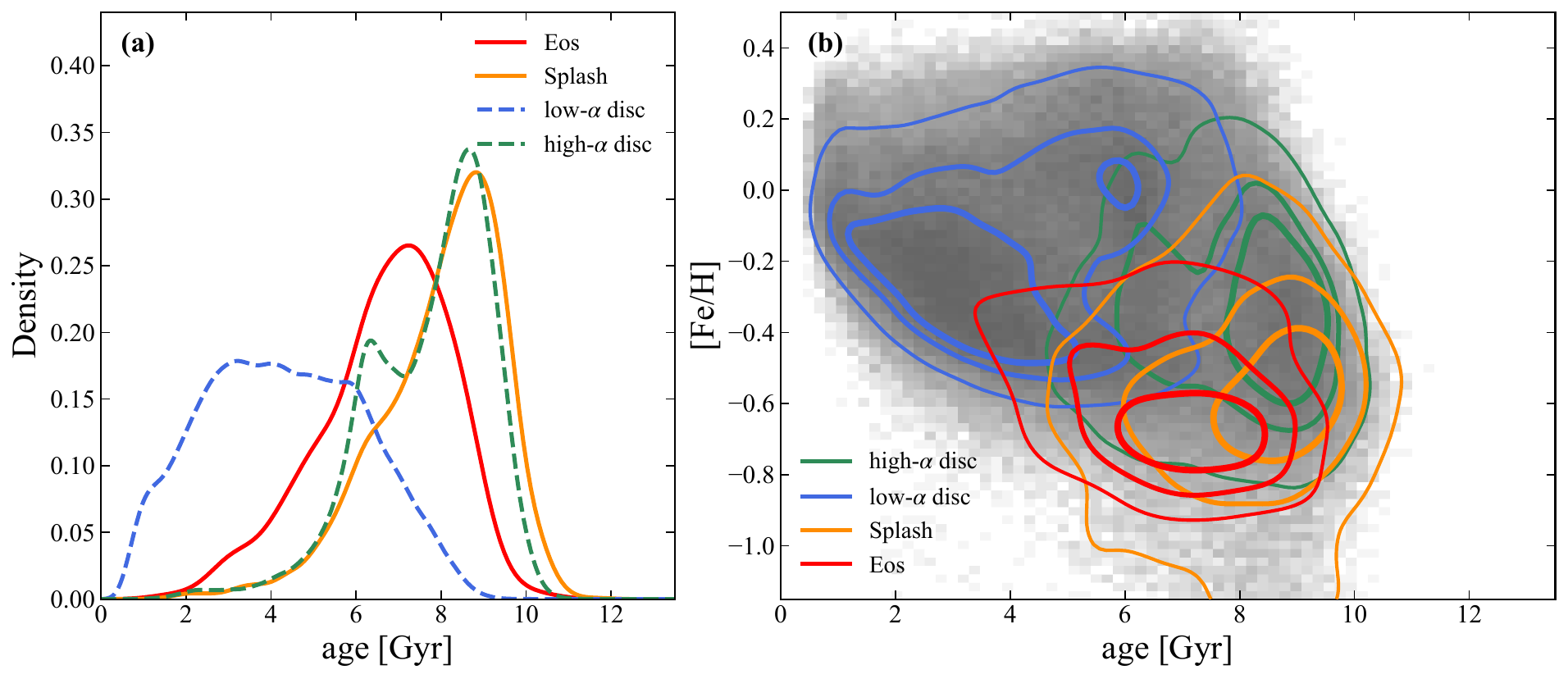}
    \caption{Ages and age--metallicity relation of Eos and the Milky Way populations, using astroNN ages. Age distributions (\textit{left}) and age--[Fe/H] distributions (\textit{right}) of Eos (red), the high-$\alpha$ disc (green), the low-$\alpha$ disc (blue) and the Splash (orange), with contours enclosing 30, 60 and 90 per cent of the stars. The samples (astroNN ages, $\sigma_{\rm age}/{\rm age}<0.3$) are Eos ($n=318$), the Splash ($n=4837$), the low-$\alpha$ disc ($n=135\,083$) and the high-$\alpha$ disc ($n=42\,309$); the grey background in the right panel is the $\log_{10}$ number density of the whole base sample in the age--[Fe/H] plane. The age distributions in the left panel are Gaussian kernel-density estimates, each normalised to unit area, with a mode of $\approx7.2$~Gyr for Eos. Eos is older than the bulk of the low-$\alpha$ disc but younger than the high-$\alpha$ disc and the Splash, and in the age--metallicity plane it extends the low-$\alpha$ sequence to older ages and lower metallicities.}
    \label{fig:amr}
\end{figure*}

Having established that Eos is a chemokinematically distinct structure from the low-$\alpha$ disc, we now investigate its ages. We adopt the spectroscopic ages from astroNN \citep[][]{Leung2019} and apply a quality cut to limit the sample to stars with relative age errors smaller than 30 per cent, i.e. $\sigma_{\rm age}/{\rm age}<0.3$ (318 of the 353 Eos stars pass this cut). The age distribution of Eos is shown by the red line in the left panel of Figure~\ref{fig:amr}. For comparison, the age distributions of the high- and low-$\alpha$ discs and the Splash are shown in green, blue and orange, respectively. The age distributions of the Splash and the high-$\alpha$ thick disc are very similar because these stars share the same origin. The mode of the Eos age distribution is intermediate between those of the high- and low-$\alpha$ discs: Eos is younger than the high-$\alpha$ disc and the Splash, but older than the low-$\alpha$ disc. This implies that Eos formed around the time of the transition between the high- and low-$\alpha$ discs, close to the time of the GS/E merger. Note that we use these ages only in a relative sense, because the absolute scale of the astroNN ages is uncertain and they lose resolution for stars older than $\approx10$~Gyr. To verify that this result is not specific to the astroNN age catalogue, we also inspect the age distributions of these stellar populations using the \texttt{BINGO} spectroscopic age catalogue of APOGEE \citep[][]{Ciuca2021} in Figure~\ref{fig:bingo_age}. The relative chronological ordering of the Splash, Eos and the low-$\alpha$ disc is preserved. The mode of the Eos age distribution shifts from $\approx7.2$~Gyr (astroNN) to $\approx12$~Gyr (\texttt{BINGO}). The mode of the Eos age in the \texttt{BINGO} catalogue is consistent with the time of the GS/E merger inferred by \citet{Ciuca2024} from a rapid chemical enrichment event, which they dubbed the Great Galactic Starburst (GGS). However, this result also reflects the inconsistency between different age catalogues. A careful calibration of the different age scales is important to set an absolute age scale for Galactic evolution. 

In the right panel of Figure~\ref{fig:amr}, we further explore the age--metallicity signature of Eos and its connection to the other stellar populations of the Milky Way. The age--metallicity contours of the high- and low-$\alpha$ discs, Eos and the Splash are shown in the same colours as before. The high-$\alpha$ disc (green) shows a steep rise in metallicity over a short timescale, illustrating the rapid chemical enrichment during the early phase of disc formation. The Splash (orange) has similar ages, but is slightly more metal-poor than the thick disc. Given the tight age--metallicity correlation observed in the high-$\alpha$ thick disc \citep[][]{Xiang_Rix2022}, the Splash population may in fact be slightly older than the high-$\alpha$ disc \citep{Sun2026}. The low-$\alpha$ thin disc is characterised by a ``V''-shaped age--metallicity sequence with a wide metallicity distribution at each age. The wide metallicity distribution in this solar-neighbourhood sample is caused by the radial mixing of stars after their birth \citep[e.g.][]{Roskar2008, Haywood2008}. The characteristic ``V'' shape of the age--metallicity sequence was discussed by \citet{Zhang2025}. They suggested that the metal-rich branch ($[\rm Fe/H]\gtrsim0$, with flat $[\rm Fe/H]$ as age decreases) consists of inner-Galaxy stars that migrated outwards, driven by the deceleration of the bar. The metal-poor branch ($[\rm Fe/H]\lesssim0$, with metallicity increasing as age decreases) consists of stars that formed locally in the outer disc. The rising metallicity along this lower branch reflects local chemical enrichment around the solar radius. Compared with the lower branch of the low-$\alpha$ disc's age--metallicity sequence, the Eos population is more metal-poor and older, and appears to be an old, metal-poor extension of the local age--metallicity sequence. This, together with its chemical connection to the low-$\alpha$ disc, suggests that the Eos population may mark the onset of star formation in the outer low-$\alpha$ disc.  

\subsection{A bifurcation in the [Fe/H]--[Mg/Fe] plane}
\label{sec:eos_bifurcation}

\begin{figure*}
    \centering
    \includegraphics[width=\linewidth]{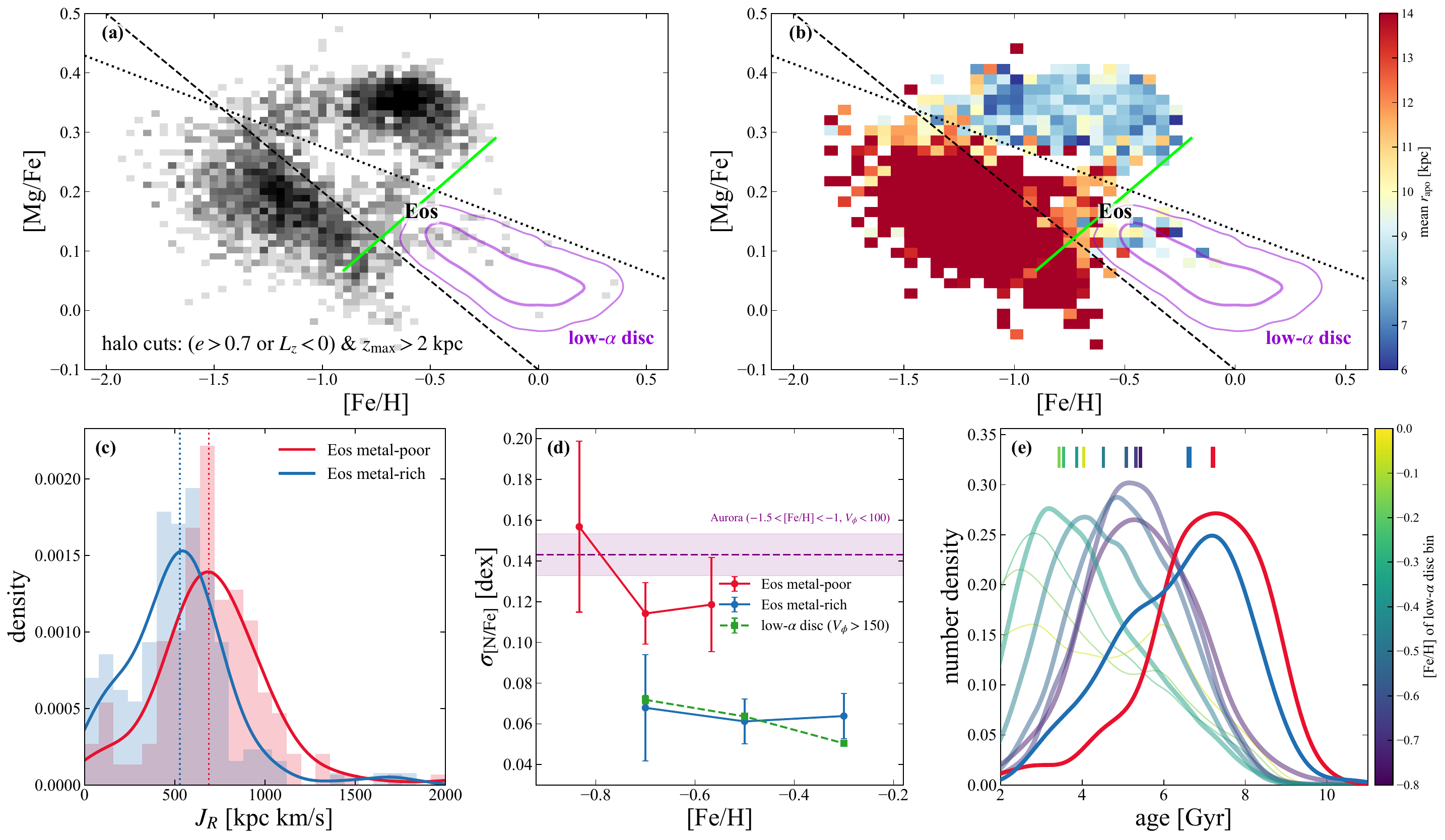}
    \caption{The two branches of Eos and their distinct properties. \textit{(a)} Density of halo-orbit stars ($e>0.7$ or $L_z<0$, with $z_{\rm max}>2$~kpc) in the [Fe/H]--[Mg/Fe] plane, with the accreted/in-situ (dashed), high-$\alpha$/low-$\alpha$ (dotted) and \citet{Davies2025} Eos-branch divider (green) lines; the dark-violet contours (50 and 90 per cent) show the low-$\alpha$ disc. \textit{(b)} The same plane coloured by the mean apocentric radius $r_{\rm apo}$. \textit{(c)} The radial-action $J_R$ distributions, \textit{(d)} the error-deconvolved nitrogen dispersion and \textit{(e)} the astroNN age distributions of the metal-poor (red) and metal-rich (blue) Eos branches. In (d), the low-$\alpha$ disc (green) and the Aurora benchmark (purple band) are shown for comparison, and in (e) the low-$\alpha$ disc is coloured by its metallicity. The metal-poor branch of Eos has larger $r_{\rm apo}$ and $J_R$ and an Aurora-like nitrogen dispersion, whereas the metal-rich branch has tighter, Splash-like orbits and a low, disc-like dispersion, indicating different formation environments for the two branches. Panel (a) uses halo-orbit stars with $z_{\rm max}>2$~kpc ($n=5372$); the green divider of \citet{Davies2025} is $[\rm Mg/Fe]=0.317\,[\rm Fe/H]+0.353$. Eos splits into $191$ metal-poor and $162$ metal-rich stars ($166$ and $152$ with astroNN ages). In (c) the dotted lines are medians. In (d) the dispersion is error-deconvolved, $\sigma=\sqrt{{\rm MAD}^2-\langle{\rm err}^2\rangle}$, with $16$--$84$th-percentile bands from $400$ bootstrap resamples and the Aurora value $0.143$. In (e) each curve is a Gaussian kernel-density estimate of one $0.1$-dex [Fe/H] bin of the low-$\alpha$ disc (thick lines for bins in the Eos range $-0.8\le[\rm Fe/H]\le-0.3$, thin outside), the ticks marking each curve's median age.}
    \label{fig:bifurcation}
\end{figure*}

\citet{Davies2025} re-analysed the halo substructures with non-negative matrix factorisation and, in their analysis of the [Fe/H]--[Mg/Fe] distribution of halo stars, noticed that Eos consists of two distinct branches in this plane with slightly different radial-action distributions. In Figure~\ref{fig:bifurcation}, we revisit this bifurcation of Eos and further explore its chemodynamical properties.

Panel (a) of Figure~\ref{fig:bifurcation} presents the [Fe/H]--[Mg/Fe] distribution of Milky Way stars with halo-like kinematics. We apply the same cuts as \citet{Davies2025} to select stars with $e>0.7$ or $L_z<0$\,kpc\,\kms, plus an extra cut on the maximum vertical height, $z_{\rm max}>2~\rm kpc$. This last cut removes bar/bulge contamination from the halo sample, predominantly stars with super-solar metallicity. Three main clouds remain in the [Fe/H]--[Mg/Fe] plane. Stars below the dashed line are the accreted GS/E population, the same as the low-[Al/Fe] cloud in the right panel of Figure~\ref{fig:mg_al}. The high-[Mg/Fe] cloud above the dotted line consists of the pre-disc Aurora and the splashed high-$\alpha$ disc stars. The Eos stars lie between the two black lines. They bifurcate into two [Fe/H]--[Mg/Fe] branches, separated by the green line. We refer to the stars on its low- and high-[Fe/H] sides as the metal-poor and metal-rich Eos stars. 

We now investigate the chemodynamics of these two branches in detail. Panel (b) of Figure~\ref{fig:bifurcation} reveals that the three main clouds in the halo's [Fe/H]--[Mg/Fe] plane have different mean apocentric radii. The high-$\alpha$ Aurora and Splash cloud has a mean apocentre of $\approx8$--$9$~kpc, far smaller than that of the accreted GS/E population, with $\langle  r_{\rm apo}\rangle \gtrsim 14 ~\rm kpc$. In comparison, the two bifurcated Eos populations have different apocentres. The metal-poor Eos stars have $\langle  r_{\rm apo}\rangle \approx11$--$12$~kpc, whereas the metal-rich Eos stars have $\langle  r_{\rm apo}\rangle \approx9 ~\rm kpc$, similar to the splashed high-$\alpha$ disc population. Panel (c) shows the radial-action distributions of the two Eos populations: the metal-poor Eos has larger radial actions than the metal-rich one, as found by \citet{Davies2025}. These clear dynamical differences hint that the metal-poor and metal-rich Eos populations might have different origins. 

Panel (d) of Figure~\ref{fig:bifurcation} further compares the nitrogen dispersion of the two Eos populations. The metal-rich Eos, shown by the blue line, displays a much lower nitrogen dispersion ($\sigma_{\rm[N/Fe]}\approx0.06$) than we found earlier for the non-rotating low-$\alpha$ stars at this metallicity ($\sigma_{\rm[N/Fe]}\approx0.10$--$0.15$; Figure~\ref{fig:N_dispersion}), a selection that mixes both branches. Its nitrogen dispersion is very similar to that of the low-$\alpha$ disc stars selected with $v_\phi>150$\,\kms, shown by the green dashed line. \new{In contrast, the metal-poor branch (red line) has $\sigma_{\rm[N/Fe]}\approx0.11$--$0.16$, about 1.6 times the disc value at [Fe/H]$\approx-0.7$ and comparable to that of Aurora. The nitrogen excess in Figure~\ref{fig:N_dispersion} is therefore carried by the metal-poor branch. The metal-rich branch has the dispersion of the low-$\alpha$ disc and Splash-like apocentres, which is what a heated disc would look like, since heating does not change the nitrogen dispersion (Section~\ref{sec:eos_nitrogen}). The two branches thus likely have different origins: the metal-poor branch formed in a stochastic, poorly mixed environment like that of Aurora, while the metal-rich branch is consistent with a disc-like birth environment.} 

Interestingly, the metal-rich Eos occupies the same part of the [Fe/H]--[Mg/Fe] plane as the most metal-poor low-$\alpha$ disc, implying a connection between these stars and the onset of star formation in the low-$\alpha$ disc. Panel (e) of Figure~\ref{fig:bifurcation} explores the ages of these two populations and compares them with the age distributions of the low-$\alpha$ disc. The mode of the age distribution of the metal-poor Eos (red) is similar to that of the metal-rich Eos (blue), but the metal-rich Eos has a more extended tail towards young ages, which makes its median age slightly younger than that of the metal-poor branch. The other age distributions in the figure correspond to low-$\alpha$ disc stars at different metallicities. While the metal-rich branch of Eos occupies the same part of chemical space as the oldest, most metal-poor low-$\alpha$ disc, it is older than the low-$\alpha$ disc stars at that metallicity (although their age distributions overlap). 

These observed properties of the two branches of Eos suggest that the metal-poor branch, with its larger orbital extent, formed in a more stochastic environment, and the metal-rich branch in a less turbulent environment like the low-$\alpha$ disc. However, we are still unable to draw a firm conclusion about the exact origin of the bifurcation, especially the origin of the metal-rich branch. The metal-rich branch may have formed in the same halo starburst as the metal-poor branch, from the partially mixed gas of the GS/E and the Milky Way (see Section~\ref{sec:au18_lane}), but closer to the Milky Way, and hence in a more Milky Way-like environment. Alternatively, the metal-rich branch may consist of early low-$\alpha$ disc stars that formed during the GS/E merger but were splashed soon after their birth (see Appendix~\ref{appendix:dual_origin}). The main difference between these two scenarios is whether the metal-rich branch first formed on disc orbits or on halo orbits, and a direct implication of this difference is whether the low-$\alpha$ disc formed during or after the GS/E merger. The dominant Milky Way population coeval with Eos is the high-$\alpha$ disc (see Figure~8 in \citealt{funakoshi2026}), which suggests that the starburst in the Milky Way's disc mainly produced high-$\alpha$ stars. However, some low-$\alpha$ stars may have formed during the same period outside the main star-forming disc, from gas accreted from the GS/E or the circumgalactic medium (as in the misaligned outer disc scenario in the Vintergatan simulation; \citealt{Renaud2021_VIII}). More detailed abundances of heavier elements and more precise ages would allow us to compare the low-$\alpha$ disc stars with the signatures of other chemical enrichment channels, such as the r-process \citep[e.g.][]{Monty2024, Kane2026b}.


\section{Insight from simulations}
\label{sec:sims}

We have discussed the observational chemodynamical signatures of this non-rotating, low-$\alpha$ disc-like population, Eos. We found that the majority of the Eos population shows a larger nitrogen dispersion than the low-$\alpha$ disc population, implying a different formation environment. In this section, we look for the origin of Eos by investigating Eos-like structures in cosmological simulations. Specifically, we explore merger-induced star formation in the Auriga cosmological simulation suite \citep[][]{Grand2017}. 

\subsection{The Auriga simulation suite, halo 18}
\label{sec:intro_au18}

\begin{figure}
    \centering
    \includegraphics[width=\linewidth]{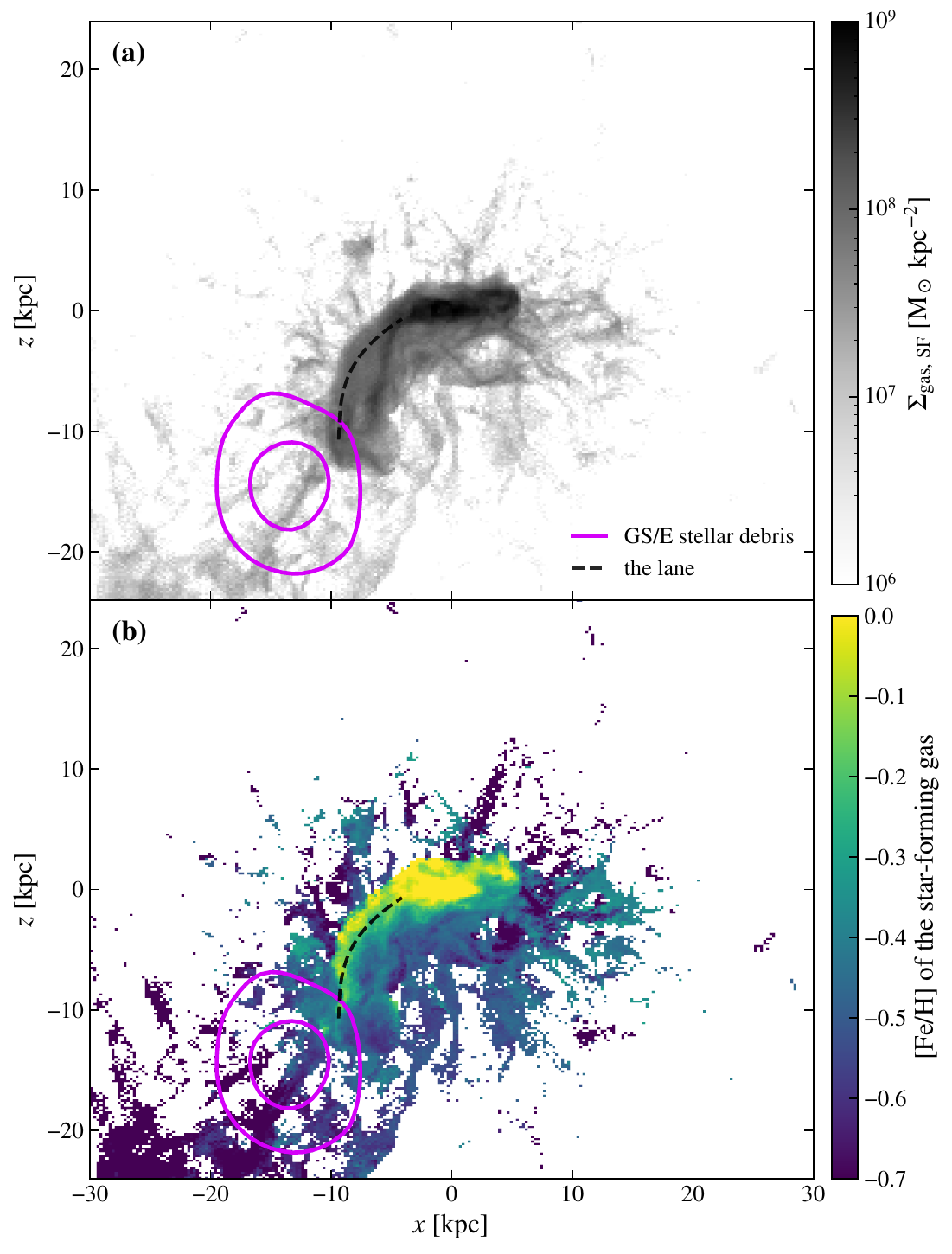}
    \caption{Edge-on maps of the star-forming gas in Au18 at the last pericentric passage of the GS/E-like merger ($t\approx5.0$~Gyr). \textit{Top:} Surface density of the star-forming gas; the magenta contours mark the stellar debris of the GS/E progenitor and the black dashed line denotes the gas lane connecting it to the main host. \textit{Bottom:} Mass-weighted mean [Fe/H] of the star-forming gas. The lane has a metallicity intermediate between those of the metal-poor GS/E progenitor and the metal-rich host, showing that stripped progenitor gas mixes with host gas along the lane. The maps are centred on the host, so its centre lies at the origin.}
    \label{fig:au18_gas}
\end{figure}

To interpret the observed chemodynamics of \Eos, we compare it with the Auriga suite of cosmological magneto-hydrodynamical zoom-in simulations \citep{Grand2017}, run with the moving-mesh code \textsc{arepo} \citep{Springel2010, Pakmor2016}. Auriga models galaxies spanning virial masses $M_{200}=1$--$2\times10^{12}\,\Msun$, evolved from
$z=127$ to $z=0$ with a \citet{Planck2014} cosmology ($\Omega_m=0.307$,
$\Omega_\Lambda=0.693$, $\Omega_b=0.048$, $h=0.677$). The galaxy-formation model includes primordial and metal-line cooling, a spatially uniform ultraviolet background that completes reionisation at $z=6$, a subgrid multiphase model for the interstellar medium, star formation, stellar evolution and feedback, magnetic fields, and black-hole seeding, accretion and feedback (see \citealt{Grand2017} for details). At the resolution level used here (level~4), dark-matter particles have masses of $\approx4\times10^{5}\,\Msun$ and gas and star particles $\approx5\times10^{4}\,\Msun$, with a softening length of $500\,h^{-1}$\,pc at $z>1$ that decreases to $369$\,pc thereafter.

\begin{figure*}
    \centering
    \includegraphics[width=\linewidth]{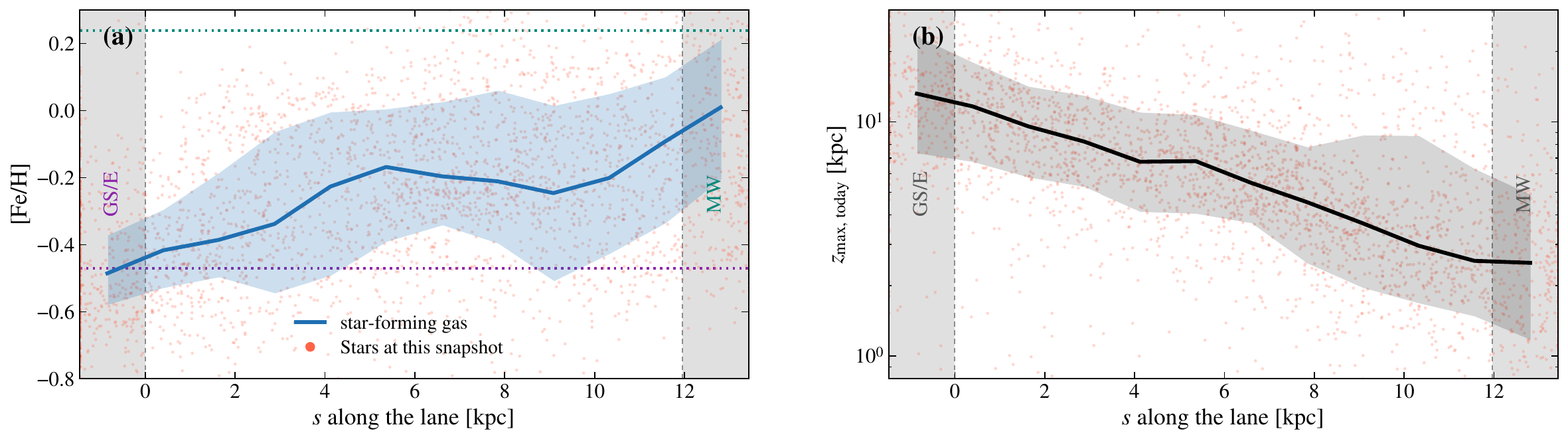}
    \caption{Metallicity (\textit{left}) and $z_{\rm max}$ (\textit{right}) of stars born along the gas lane in Au18 as a function of position along the lane, $s$ ($s\approx0$ near the GS/E progenitor, increasing towards the host). \textit{Left:} points show individual newly born stars; the blue line and band show the running mean and 16--84th percentile range of the star-forming gas; and the dotted lines show the mean metallicities near the GS/E progenitor (purple) and the host (teal). \textit{Right:} Present-day maximum vertical height $z_{\rm max}$ of the same stars, with the median (black) and 16--84th percentile band. Metallicity rises overall as the gas mixes along the lane, and birth position maps onto a present-day $z_{\rm max}$ gradient, so that stars born nearer the GS/E progenitor settle on more extended orbits. The plotted stars are those born in the $\approx150$~Myr before the snapshot within a $3$~kpc tube around the lane, and $s$ is measured by projecting each star onto the lane centreline.}
    \label{fig:au18_lane_fehzmax}
\end{figure*}

We focus on a single halo, Au18. \citet{Fattahi2019} showed that the stellar halo of Au18 has a strong radial anisotropy produced by a GS/E-like merger that occurred $\approx9$\,Gyr ago (cosmic time $t\approx5$\,Gyr), closely mirroring the accretion history inferred for the Milky Way. The progenitor reaches apocentre at $t\approx3.25$~Gyr and its last pericentre at $t\approx5.0$~Gyr, and coalesces at $t\approx5.3$--$5.6$~Gyr, bringing in $\approx10^{10}~\Msun$ of gas \citep{Grand2020}. Since this work concerns the response of the in-situ Milky Way stars to the GS/E merger, Au18 is particularly well suited to our purpose; moreover, like the Milky Way, Au18 develops both a high-$\alpha$ and a low-$\alpha$ disc \citep{Orkney2026a_bimodality}. We stress, however, that no cosmological galaxy is a perfect analogue of the Milky Way; we therefore use Au18 to identify the physical mechanisms at play rather than to reproduce the Galaxy in detail.

Because the chemodynamical properties of the Eos stars suggest a strong connection to the GS/E merger, we zoom into the last pericentric passage of the GS/E-like merger in Au18 to understand in detail how stars form during such a merger. Au18 has been studied extensively in the literature \citep[e.g.][]{Grand2020, Ciuca2024, Orkney2026a_bimodality}, particularly with regard to the impact of the GS/E-like merger on the Milky Way-analogue galaxy. \citet{Grand2020} and \citet{Ciuca2024} showed that a star-formation burst occurred at the last pericentric passage of the GS/E progenitor in Au18, and a similar burst was described in detail for a high-resolution re-run of Au24 by \citet{Orkney2026b_}. Below, we revisit some of their results to place the Eos population discussed above in context, focusing on the chemodynamics of the stars formed in this burst at the last pericentric passage of the GS/E progenitor. 

\subsection{Stars formed along the gas lane}
\label{sec:au18_lane}

\begin{figure*}
    \centering
    \includegraphics[width=\linewidth]{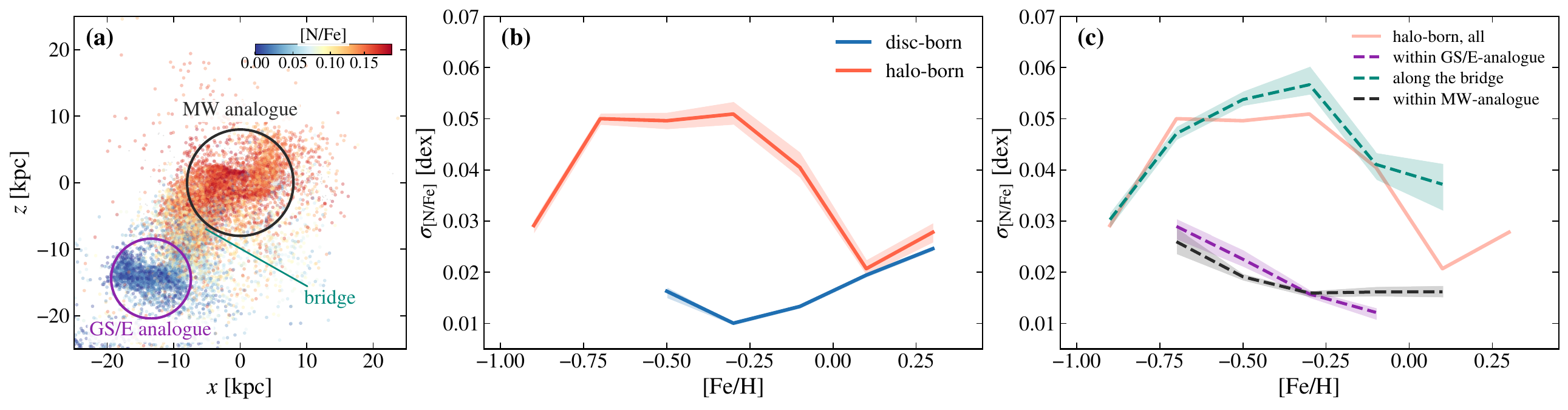}
    \caption{Chemical dispersion of stars formed during the GS/E-like merger in Au18. \textit{Left:} Edge-on map of newborn stars coloured by [N/Fe]; the circles mark the apertures defining stars born within the GS/E analogue (purple) and within the host (dark), and the green label marks the bridge/lane. \textit{Middle:} The nitrogen dispersion $\sigma_{\rm [N/Fe]}$ versus metallicity for halo-born (red) and disc-born (blue) stars, computed as $1.4826\times$MAD in $0.2$-dex bins, with bands showing the $16$--$84$th percentiles of $500$ bootstrap resamples; bins with fewer than $150$ stars are not drawn. \textit{Right:} The halo-born dispersion split by birth site: within the GS/E analogue (purple), along the bridge (green), and within the host (black). At low metallicity, halo-born stars have a larger dispersion than disc-born stars, and the lane-born stars have the largest dispersion of all because they sample the least-mixed gas; this reproduces the enhanced dispersion observed in Eos.}
    \label{fig:au18_Ndisp}
\end{figure*}

We rotate Au18 so that the net angular momentum vector of the disc is aligned with the z-axis. The upper panel of Figure~\ref{fig:au18_gas} shows the surface density of star-forming (SF) gas in the edge-on view of the galaxy at the last pericentric passage of the GS/E-like merger. The magenta contours show the stellar density of the GS/E progenitor and mark its location at this snapshot. As in Figure~1 of \citet{Grand2020}, a gas lane, denoted by the black dashed line, connects the GS/E progenitor to the main host galaxy. No prominent gas cloud remains inside the inner contour of the GS/E progenitor, which means that its gas has been stripped and is being accreted onto the main host, mixing along the lane with gas outflowing from the host. This is why the gas lane has a metallicity intermediate between that of the gas close to the GS/E progenitor ([Fe/H]$_{\rm gas}\approx-0.5$ to $-0.3$) and that of the gas close to the main host ([Fe/H]$_{\rm gas}\gtrsim0$), as seen in the lower panel of Figure~\ref{fig:au18_gas}, which shows the mean metallicity of the SF gas. The left panel of Figure~\ref{fig:au18_lane_fehzmax} illustrates this metallicity transition more clearly. There, we show the metallicity along the gas lane, marked by the black dashed line in Figure~\ref{fig:au18_gas}. We compute the distance along the lane, $s$, where $s\approx0$ is close to the GS/E progenitor and $s\approx12~\rm kpc$ is close to the main host. The red dots in the background show the metallicities and positions along the lane of the newly born stars (those formed in the $\approx150$~Myr before this snapshot) at the GS/E's last pericentric passage ($t_{\rm snap}=5.0~\rm Gyr$). The blue line and shaded region show the running mean metallicity of the SF gas and its 16--84th percentile range. The metallicity rises overall along the lane from the GS/E progenitor to the main host, whose mean metallicities are shown by the purple and teal dotted lines, respectively. This implies that the gas mixes rapidly along the lane during the merger.

The chemical properties of stars born along the lane are explored further in Figure~\ref{fig:au18_Ndisp}. The left panel shows all stars born within $150~\rm Myr$ before this snapshot, coloured by nitrogen abundance. The purple and black circles mark the locations of the GS/E progenitor and the main host. The nitrogen abundance also shows a clear gradient in the left panel: [N/Fe] is low in the GS/E progenitor and high in the main host. As with metallicity, stars along the gas lane have intermediate [N/Fe] because of gas mixing. As we demonstrated in Section~\ref{sec:eos_nitrogen}, the Eos population has a large dispersion in elements such as nitrogen, with a $95\%-5\%$ percentile range similar to that of the pre-disc Milky Way. Here, in the middle and right panels of Figure~\ref{fig:au18_Ndisp}, we inspect the elemental dispersion of stars formed along the lane in Au18 as a function of metallicity. In the middle panel, we compare the nitrogen dispersion of stars born on halo-like and disc-like orbits. Halo-like orbits are defined by orbital circularity $\epsilon<0.8$ ($\epsilon=L_z/L_{\rm circ}(E)$) and maximum vertical height of the birth orbit $z_{\rm max,\,birth}>1.5~\rm kpc$, while all other stars are classed as being on disc-like orbits. The blue line shows the chemical dispersion of the disc-born stars and the red line that of the halo-born stars, which predominantly formed along the lane. The chemical dispersion of the halo-born stars is higher than that of the disc-born stars at low metallicity ([Fe/H]$\lesssim-0.2$). We further split the halo-born stars into three groups: stars formed within the main host ($|\bs{r}|<8~\rm kpc$, the black circle in the left panel), stars formed within the GS/E progenitor ($|\bs{r}-\bs{r_{\rm GS/E}}|<6~\rm kpc$, the purple circle), and stars formed along the gas lane (the halo-born stars lying outside both apertures). The right panel of Figure~\ref{fig:au18_Ndisp} shows the nitrogen dispersion of each group. Stars formed along the gas lane (green line) have a larger dispersion than stars formed around the main host and the GS/E progenitor. This is because the gas in the lane is not yet fully mixed and so has a large abundance dispersion; stars formed along the lane sample this inhomogeneous gas and inherit its large chemical dispersion. The gas is better mixed closer to the main host and the GS/E progenitor. This also explains the drop in chemical dispersion of the halo-born stars at high metallicity: stars formed near the main host dominate these bins, and the larger their contribution, the lower the dispersion.
We show the nitrogen dispersion here only for consistency with the observations, but we see the same high dispersion in other abundances, such as Mg and Si, so this result does not depend on the nitrogen yields adopted in the Auriga simulations. Overall, the chemical properties of stars formed in the gas lane closely match those of the Eos population.

Stars inherit the velocity of the gas from which they form. The lane consists of gas stripped from the GS/E progenitor, which at its last pericentre moves on a nearly radial, low-angular-momentum orbit, mixed with gas flowing out of the host. Stars born in the lane therefore start with little net rotation and large radial velocities, and their orbital energies lie between those of the GS/E debris and the host. Stars born closer to the GS/E progenitor start further out and move more like it, producing the gradient seen in the right panel of Figure~\ref{fig:au18_lane_fehzmax}.

The dynamical properties of stars born along the lane also resemble those observed for Eos. Because stars born at different places along the gas lane have different integrals of motion, we expect a gradient in their dynamical properties. The right panel of Figure~\ref{fig:au18_lane_fehzmax} shows this gradient clearly, relating $z_{\rm max,\,today}$ of each star to its birth site along the gas lane. Stars born closer to the GS/E progenitor have a larger $z_{\rm max}$ than those born closer to the host. We showed above that the chemistry of stars formed along the lane changes from GS/E-like to host-like, depending on where along the lane the stars formed. Therefore, the correlation between $z_{\rm max}$ and $s$ should appear as a relation between chemistry and dynamics: stars with chemistry similar to that of the GS/E progenitor (in Au18, lower [Fe/H]) have larger $z_{\rm max}$, and stars with chemistry similar to that of the Milky Way have smaller $z_{\rm max}$. This is what we observed in Figure~\ref{fig:al_fe_orb}: the Eos stars with low [Al/Fe] (similar to the GS/E chemistry) have larger $z_{\rm max}$ and $r_{\rm apo}$, and those with high [Al/Fe] have smaller $z_{\rm max}$ and $r_{\rm apo}$. Stars born along the gas lane during the merger, from a mixture of gas from the GS/E progenitor and the Milky Way analogue, show chemodynamical properties similar to those of the observed Eos stars, suggesting a possible birth environment for Eos. 

\subsection{Star-formation burst in the halo}
\label{sec:sim_sfr}

\begin{figure*}
    \centering
    \includegraphics[width=\linewidth]{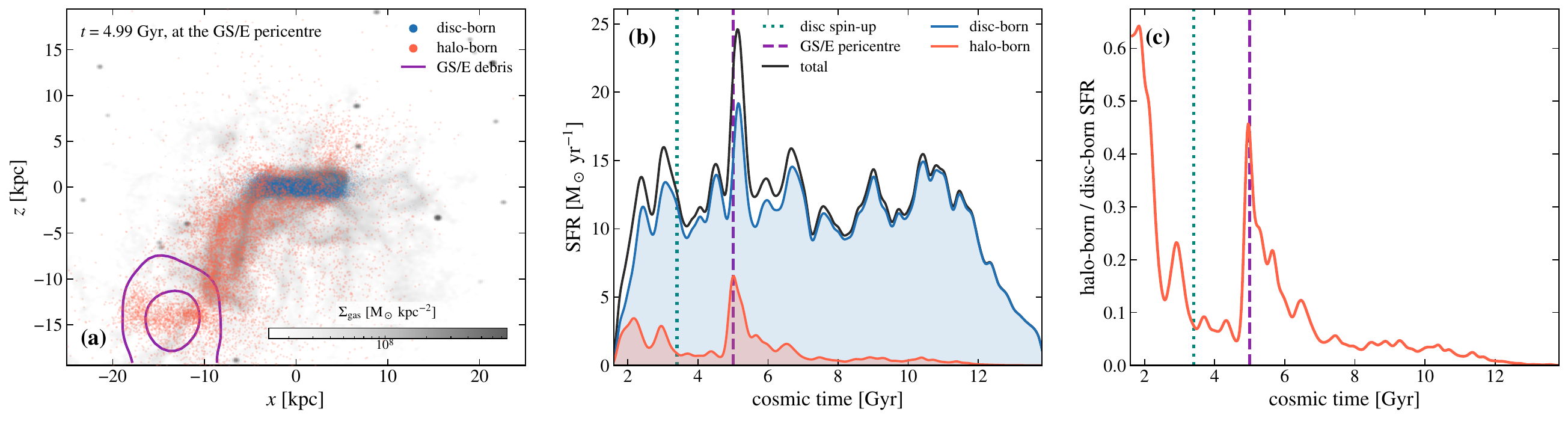}
    \caption{Star formation on halo and disc orbits during the GS/E-like merger in Au18. \textit{Left:} Edge-on gas surface-density map at the last pericentric passage ($t\approx5.0$~Gyr) with the positions of newly born disc-orbit (blue) and halo-orbit (red) stars; the purple contours mark the GS/E stellar debris. \textit{Middle:} The disc-born (blue), halo-born (red) and total (black) star-formation histories versus cosmic time; the dotted and dashed vertical lines mark the disc spin-up (taken as the time the halo-to-disc birth-rate ratio falls below $\approx0.1$, $t\approx3.4$~Gyr) and the GS/E pericentre. \textit{Right:} The ratio of halo-born to disc-born star-formation rate. All masses and rates use birth (initial) stellar mass. Halo-orbit star formation peaks sharply at the merger, reaching $\approx30$ per cent of all star formation at the peak and $\approx20$ per cent averaged over the following $500$~Myr, producing a substantial, chemically intermediate population, formed largely along the gas lane.}
    \label{fig:au18_sfr}
\end{figure*}

Having linked the Eos stars to the merger-induced halo starburst, particularly along the gas lane, we now inspect the star-formation rate during the last pericentric passage of the GS/E progenitor. The left panel of Figure~\ref{fig:au18_sfr} shows the spatial distribution of stars born within 150~Myr before this snapshot, separated into stars born on disc orbits and on halo orbits using the same criteria on birth orbital circularity and $z_{\rm max}$. The spatial distribution of the selected stars shows that this classification works well: as seen in the left panel of Figure~\ref{fig:au18_sfr}, the blue stars follow a flattened disc morphology, while the red stars are halo-like. A significant number of stars do not form on disc orbits during the merger, as also found by \citet{Grand2020, Orkney2026b_}. Of the stars formed on halo orbits, a significant fraction ($\sim60$ per cent) lie close to the gas lane connecting the main host and the GS/E progenitor. The other halo-born stars sit either with the undisrupted GS/E progenitor or around the main host. 


The star-formation rate (SFR) of Au18 is shown by the black solid line in the middle panel of Figure~\ref{fig:au18_sfr}. It peaks around the last pericentric passage of the merger, which corresponds to the merger-induced starburst \citep[][]{Grand2020, Ciuca2024, Laporte2026}. We also show the SFR of stars born on disc and halo orbits as blue and red lines, respectively. Although the peak appears in both populations, the burst is significantly more pronounced, relative to other times, for stars on halo orbits. This is illustrated further in the right panel, which shows the ratio of halo-born to disc-born SFR. While the ratio is below $\approx0.1$ for most of the time after the disc spin-up, it peaks around the last pericentric passage, reaching $\approx0.46$. Around the time of the merger (within 500~Myr after the last pericentric passage), $\approx20$--$30$ per cent of stars formed on halo orbits. These halo-born stars have a total birth mass of $\approx2.9\times10^9~\rm M_{\odot}$, read from the halo-born star-formation history in Figure~\ref{fig:au18_sfr} over this window; stars formed along the gas lane account for about half of this mass, and stars formed around the GS/E progenitor for only 5 per cent. If we select stars born on disc orbits between the disc spin-up and the GS/E merger, but with present-day $|v_\phi|<80$\,\kms (the Splash in Au18), the Splash has a birth mass of $\approx3.1\times10^{9}~\rm M_{\odot}$, comparable to the mass of this halo-born population and to that of the stars formed along the gas lane. This implies that we could expect to find the population born on halo orbits at the time of the last pericentric passage of the GS/E as part of the Milky Way's in-situ halo, and this population is likely to be the Eos population discussed above. Although \citet{Orkney2026b_} argued the contrary, namely that the chemically distinct, merger-induced halo-born population might be difficult to detect because of rapid gas mixing, we find in Au18 that the stars formed in the gas lane have chemistry intermediate between the two galaxies and could contribute a substantial fraction of the mass of the local in-situ stellar halo. The GS/E-like merger in Au18 has an accompanying galaxy falling at the same time with similar orbits could also contributed some of the stellar masses formed during this starburst, which might be a reason why we measured a more massive Eos-like component in Au18 than Au24 rerun as in \citet{Orkney2026b_}.

The mass of stars formed in the gas lane in Au18 is more than 10 times higher than the estimated mass of the Eos population in the Milky Way (see below). \citet{Myeong2022} fitted a Gaussian mixture model (GMM) in the chemical space of local halo stars with eccentricity $>0.85$ and found that Eos stars make up roughly 10--15 per cent of the stars in the local stellar halo, while the GS/E and Aurora dominate the rest of the local stellar halo \citep[][]{Zhu2024, Zhang2024}. \citet{An2025} found a population of low-$\alpha$, high-Na stars with high proper motions, similar to Eos, and derived a mass of $5\times10^{7}~\rm M_\odot$ for the Eos population by comparing the number counts of Eos and accreted GS/E stars in the local stellar halo, with a rough correction for the metallicity selection functions of different spectroscopic surveys. With the new mass estimate of GS/E from \citet{Lane2026}, which is lower than other estimates, the mass of Eos would be even lower than $5\times10^{7}~\rm M_\odot$. Following a similar approach to \citet{An2025}, we can provide another mass estimate for Eos by comparing the number counts of Eos and Aurora stars; adopting the mass of Aurora with $-1.5<[\rm Fe/H]<-1.3$ from \citet{Kurbatov2024}, we obtain $M_{\rm Eos}\sim 1\times 10^{8}~\rm M_\odot$. Because the dynamical properties of Eos are intermediate between those of the GS/E and Aurora, its actual mass should lie between the estimates anchored to the GS/E and Aurora masses, i.e. $5\times10^{7}~{\rm M_\odot}\lesssim M_{\rm Eos}\lesssim 1\times 10^{8}~\rm M_\odot$. Both limits are much lower than the mass of stars formed in the gas lane in Au18 during the merger. This could be explained either if the GS/E merger was less gas-rich than the one in Au18, so that it induced a weaker starburst, or if the gas mixed more rapidly, so that the stars formed along the gas lane are less chemically distinguishable from the GS/E and Milky Way in-situ stars, as suggested by \citet{Orkney2026b_}. Using a high-resolution re-run of Au24, \citet{Orkney2026b_} argued that the time window for producing chemically intermediate stars could be as short as a few tens of millions of years because of rapid gas mixing. Since the SFR of stars born on halo orbits could reach $\approx6.5~\rm M_\odot/yr$ during the GS/E merger (with stars born along the lane contributing about half of this SFR), the mass of stars expected to form during this short mixing window lies within the mass range we estimated for Eos.

\section{Comparison with other works}
\label{sec:comparison}

\begin{figure}
    \centering
    \includegraphics[width=0.99\linewidth]{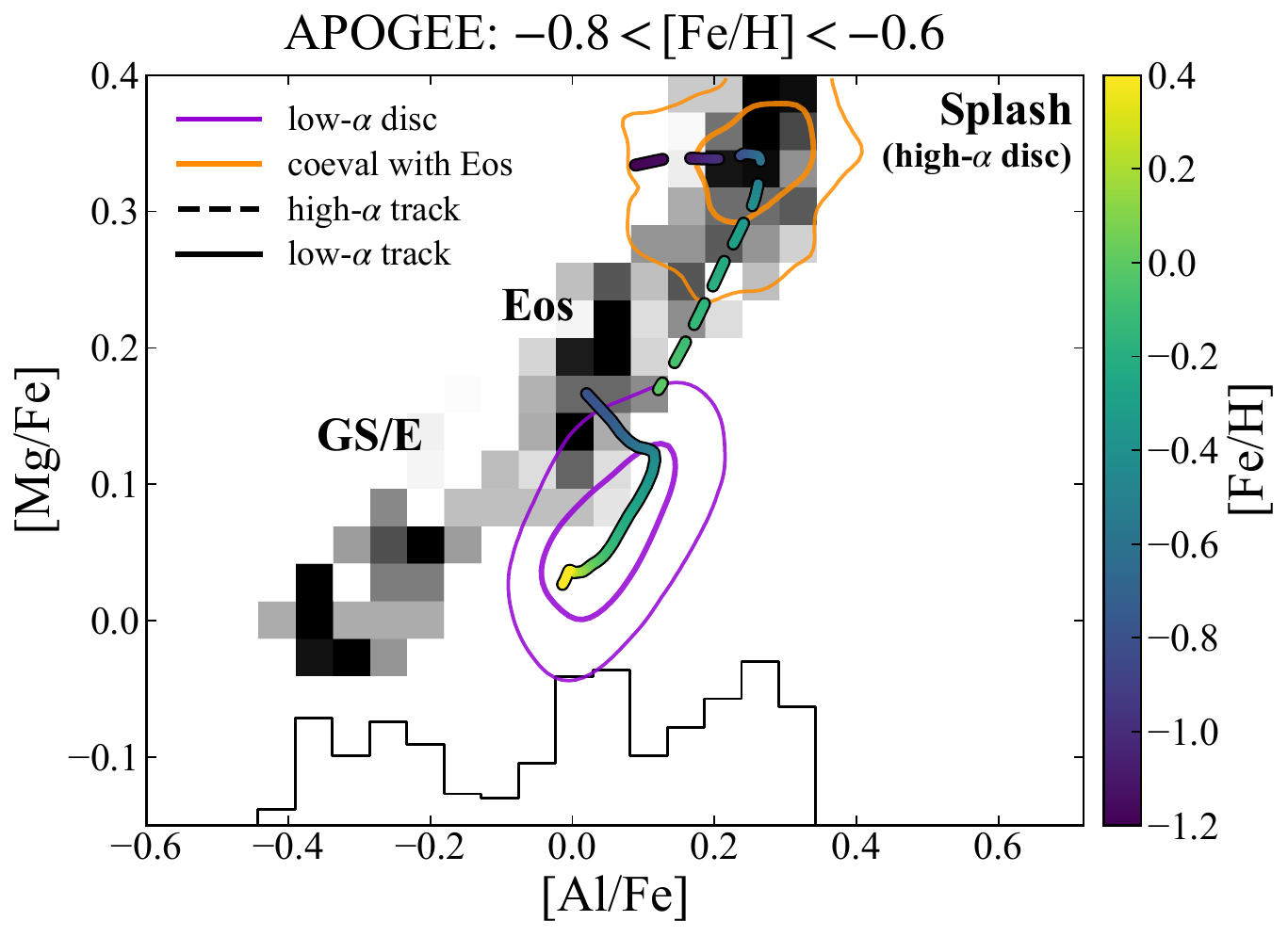}
    \caption{The intermediate chemistry of Eos in the [Al/Fe]--[Mg/Fe] plane. The grey image shows halo-orbit stars in the slice $-0.8<$[Fe/H]$<-0.6$, which form three clouds corresponding to the accreted GS/E, Eos and the Splash, as labelled. The dark-orange contours show the in-situ Milky Way population coeval with Eos and the dark-violet contours the low-$\alpha$ disc; the dashed and solid curves trace the median [Al/Fe]--[Mg/Fe] evolution of the high-$\alpha$ and low-$\alpha$ samples, coloured by metallicity. Eos lies between the GS/E debris and the coeval Milky Way population, and it connects to the chemical evolution of the low-$\alpha$ disc much as Aurora (the metal-poor end of the dashed track) connects to that of the high-$\alpha$ disc. The coeval population (orange) is selected with astroNN ages ($\sigma_{\rm age}/{\rm age}<0.3$) within $0.5$~Gyr of the median Eos age ($7.24$~Gyr), $-0.8<$[Fe/H]$<-0.6$, $-0.75\times10^5<E<-0.4\times10^5$~(km\,s$^{-1}$)$^2$ and disc orbits ($n=971$). The black histogram is the column sum of the row-normalised grey image, not the [Al/Fe] distribution of the grey stars.}
    \label{fig:mg_al_plane}
\end{figure}

In this section, we compare the observed properties and proposed origins of Eos with other studies in the literature, and extend these results to discuss what its chemodynamical properties imply for the formation of the Galactic low-$\alpha$ disc. We have presented the chemodynamical properties of this non-rotating, in-situ low-$\alpha$ population, Eos. We trace its origin to the GS/E merger-induced starburst by comparing these observed properties with the chemodynamics of stars formed during the merger in Au18, in particular the stars that formed along the gas accretion lane between the Milky Way and GS/E, where the gas was mixing. Figure~\ref{fig:mg_al_plane} summarises the chemical properties of Eos. It shows the intermediate chemical nature of Eos for halo stars in the metallicity slice $-0.8<[\rm Fe/H]<-0.6$, in the [Mg/Fe]--[Al/Fe] plane. Three distinct clouds correspond to GS/E, Eos and the Splash stars. The orange contour shows the chemistry of the Milky Way population coeval with Eos. The three populations lie roughly on one line in this plane: the GS/E debris at ([Al/Fe], [Mg/Fe])$\approx(-0.3, 0.05)$, Eos at $\approx(0.05, 0.2)$ and the coeval Milky Way stars at $\approx(0.3, 0.35)$, so that Eos lies between the accreted dwarf and its host. The purple contour shows the low-$\alpha$ disc of the Milky Way, and the dashed and solid lines show the chemical evolution tracks of the Milky Way's high- and low-$\alpha$ samples, coloured by metallicity.

\textbf{{\emph{Properties and origins of Eos:}}} The Eos population was first identified by \citet{Myeong2022}, who applied a GMM to the chemical and dynamical space of highly eccentric stars in the solar neighbourhood. Their GMM revealed Eos as a distinct chemodynamical structure that sits at the most metal-poor end of the low-$\alpha$ disc sequence in the [$\alpha$/Fe]--[Fe/H] space and has abundances of light elements such as Mg and Al intermediate between those of the Milky Way's thick disc and the GS/E. \citet{Davies2025} later applied non-negative matrix factorisation (NMF) to the chemical space and confirmed that Eos is a distinct structure. Both works suggested that the Eos stars formed from gas accreted from the GS/E. \citet{An2025} examined low-$\alpha$, high-proper-motion stars in the solar neighbourhood across many spectroscopic surveys and found a bimodal distribution in metallicity, [Na/Fe] and [Al/Fe]. The low-$\alpha$, high-[Na/Fe] (LAHN) stars in their study likely include Eos, which would form their high-eccentricity part. They suggested that these stars are the ``first sparks'' of the GS/E merger starburst and are analogues of the clumpy star formation seen at high redshift. This formation environment is similar to the one we propose in this work, in which the stars form along the gas lane between the Milky Way and the GS/E progenitor, and it would be compatible with the high nitrogen dispersion we find in the Eos population. 

The age signature of Eos is also compatible with other studies of the GS/E merger starburst. \citet{Ciuca2024} used spectroscopic ages of APOGEE stars to study the age--[Fe/H]--[$\alpha$/Fe] space and found a drop in [Fe/H] with rising [$\alpha$/Fe], followed by rapid enrichment, which they referred to as the Great Galactic Starburst (GGS) induced by the GS/E merger. The GGS happened between 12 and 10 Gyr ago on the BINGO age scale \citep[][]{Ciuca2021}. As we show in panel (b) of Figure~\ref{fig:bingo_age} in Appendix~\ref{appendix:bingo_age}, the age peak of Eos is consistent with the GGS on the same age scale. \citet{Laporte2026} also studied the timing of the GS/E merger-driven starburst using a population of globular clusters, which they dubbed \textit{Taina}, that have very similar ages but a diverse range of orbits, consistent with a starburst signature. The age of the Taina globular clusters is $\approx11.2~\rm Gyr$, which is broadly consistent with the age of Eos in the BINGO age catalogue. However, because the relative calibration of the spectroscopic ages and the isochronal ages of the globular clusters is uncertain, it is difficult to compare the age estimates of the GS/E-driven starburst quantitatively. The first-generation stars of the Taina clusters occupy a similar age--metallicity region to Eos but are more $\alpha$-enhanced \citep{Laporte2026}; thus, although the two may share the same starburst episode, the Taina clusters more likely formed within the Milky Way, while the Eos stars might have formed from partially mixed gas from the Milky Way and the GS/E progenitor. A more robust age estimate for the Eos stars could provide a good clock for the last pericentric passage of the GS/E and, when compared with the ages of the Taina globular clusters, help us better understand this starburst episode in the Milky Way. 

Other studies have proposed formation scenarios for Eos that differ from the one in this work. \citet{Orkney2026b_} argued that rapid chemical enrichment would make the Eos population very difficult to detect, and that an Eos-like population could instead form in the centre of the GS/E progenitor during the merger. This scenario is essentially the same as ours except for the exact birthplace. However, the mixing gas along the lane remains the more likely environment, because only the stars formed along the gas lane show a significantly higher chemical dispersion than the disc-born stars, as shown in Figure~\ref{fig:au18_Ndisp}. Stars formed near (or inside) the GS/E would have a chemical dispersion similar to that of the disc-born stars, which is inconsistent with the observations in Section~\ref{sec:eos_nitrogen}. The mass discrepancy between the lane stars in Au18 and Eos is addressed in Section~\ref{sec:sim_sfr}, where we show that the mass expected to form along the lane during the short mixing window is consistent with the estimated mass of Eos. Another formation mechanism compatible with the Eos stars was presented by \citet{Agertz2021_VI, Renaud2021_VIII} with the Vintergatan simulation suite. They found that star formation occurred in a spatially extended outer disc owing to tidal compression during the GS/E-like merger. \citet{Renaud2021_VIII} showed that the stars formed in this tilted outer disc have ages and $\alpha$ abundances similar to those of the metal-poor end of the low-$\alpha$ disc but halo-like kinematics, similar to the Eos stars. No results were reported on the chemical dispersion or the detailed chemodynamical relations of this population, so it is difficult to compare it with Eos quantitatively. Note that there is no essential difference between Eos forming in the gas lane and in the misaligned disc: both mechanisms form Eos in a turbulent environment, unlike that of the disc, and they differ only in the origin of the polluting gas.

\citet{Borbolato2026} proposed a distinct, merger-independent formation channel for the metal-rich (roughly $[\rm Fe/H]>-0.6$) Eos population. They analysed a subset of the GASTRO library \citep{Amarante2022, Amarante2026}, which follows the evolution of a Milky Way-like galaxy with and without an early clumpy phase and a GS/E-like merger. The high-density, star-forming clumpy phase reproduces several properties of the Milky Way disc \citep{Clarke2019, BeraldoeSilva2020, Amarante2020} and the chemistry of the bulge \citep{Debattista2023} without requiring a merger. Because of the high star-formation-rate density of the clumps, the clumpy phase produces a significant fraction of old ($>10$~Gyr), low-$\alpha$ stars \citep{BeraldoeSilva2021, Amarante2026}, similar to what has recently been observed in the Milky Way \citep[e.g.][]{Laporte2020, Borbolato2025}. \citet{Borbolato2026} focused on the counterpart of the Splash population \citep{DiMatteo2019, Belokurov2020} in the low-$\alpha$ disc. They studied low-$\alpha$ disc stars with $[\rm Fe/H]>-0.6$ and low angular momentum, $v_\phi<100~\kms$, i.e. the low-$\alpha$ Splash. Their selection therefore overlaps with the definition of the metal-rich Eos in the present work. As shown in Figure~\ref{fig:bifurcation}d, the metal-rich Eos has a $[\rm N/Fe]$ dispersion similar to that of the low-$\alpha$ disc, consistent with in-situ formation, and thus a clumpy model can also account for it; the eccentricity and $z_{\rm max}$ distributions of the low-$\alpha$ Splash are also within the expectations for the metal-rich Eos (see Figure~4 of \citealt{Borbolato2026}). \citet{Borbolato2026} removed stars formed close to the GS/E dwarf, which likely removed most stars formed in the GS/E--Milky Way lane near the last pericentre; thus, the formation of the low-$\alpha$ Splash / metal-rich Eos in the Milky Way can be attributed to in-situ bursty star formation, driven by clumps, by the merger or by both. The observational evidence presented in this paper cannot establish whether the metal-rich Eos population formed as part of the GS/E-induced starburst or is the heated, early-formed low-$\alpha$ disc population.

\textbf{\emph{Implications for the formation of the low-$\alpha$ disc:} }
Aside from its value for studying the merger-induced starburst, Eos has another interesting characteristic: its age and chemical properties are similar to those at the start of the low-$\alpha$ disc sequences. We showed in Figures~\ref{fig:3pops} and \ref{fig:bifurcation} that Eos lies at the start of, and follows, the chemical evolution trends of the low-$\alpha$ disc in the [Fe/H]--[Al/Fe] and [Fe/H]--[Mg/Fe] planes, and in Figure~\ref{fig:amr} that it also connects to the age--metallicity sequence of the locally formed low-$\alpha$ thin disc. These similarities between the GS/E merger-induced starburst stars and the oldest low-$\alpha$ thin disc stars hint at a causal connection between the GS/E merger and the formation of the low-$\alpha$ disc and the $\alpha$-bimodality. The overall picture is very similar to that in the Vintergatan simulation \citep[][]{Agertz2021_VI}, in which the stars formed in the misaligned outer disc during the merger mark the onset of low-$\alpha$ disc formation through tidal compression. The only difference is whether the gas comes mainly from the circumgalactic medium (CGM) or from a mixture of the GS/E gas and the Milky Way's outflow. Since a single GS/E-like merger has been shown to be unable to produce a long-lived metallicity dilution of more than 0.3 dex \citep[e.g.][]{Orkney2026a_bimodality}, the gas from the GS/E cannot contribute significantly to the star-forming gas of the whole low-$\alpha$ disc sequence. It more likely marks only the onset of the low-$\alpha$ thin disc and contributes only to its first and oldest population, such as the Eos stars. More recently, \citet{Genina2026} simulated and examined four Auriga-like galaxy models and stressed the importance of the merger-induced starburst and quenching in producing an $\alpha$-bimodality like that of the Milky Way. They argued that the gas accreted from the GS/E mainly feeds the starburst in the high-$\alpha$ disc, which agrees with our finding that Eos has chemistry intermediate between the GS/E and the coeval high-$\alpha$ stars in the Milky Way (see Figure~\ref{fig:mg_al_plane} and Figure~8 in \citealt{funakoshi2026}). In their models, however, the gas that forms the metal-poor low-$\alpha$ disc comes mainly from the CGM. In this scenario, it is unclear, without examining those simulated galaxies in detail, whether an Eos-like population with chemistry similar to the metal-poor low-$\alpha$ disc could still form as part of the starburst. The fact that the GS/E-driven starburst stars are only slightly older than, and chemically similar to, the metal-poor low-$\alpha$ disc stars makes them particularly interesting for studying the relation between the GS/E merger and the formation of the low-$\alpha$ disc. 

Although the emergence of $\alpha$-bimodality has been shown to be uncorrelated with the presence of a GS/E-like merger \citep[][]{Orkney2026a_bimodality}, the formation of the $\alpha$-bimodality in the Milky Way has recently been widely associated with the GS/E merger \citep[e.g.][]{Johnson2025, Amarante2026, Genina2026}. These works particularly emphasise the role of the retrograde GS/E merger. A retrograde merger drives gas inwards more efficiently than a prograde merger, which does the opposite. This quenches star formation and makes it easier for metal-poor gas from either the GS/E or the CGM to dilute the metallicity of the outer disc and form the low-$\alpha$ thin disc stars \citep[e.g.][]{Khoperskov2021}. Observationally, \citet{Funakoshi2025} found a temporary shrinkage of the disc scale length around the time of the GS/E merger by fitting action-based dynamical models to APOGEE stars with BINGO ages, which could be a signature of the inflow of the gaseous disc. Following this, \citet{funakoshi2026} linked the GS/E merger-driven starburst, the shrinkage of the gaseous disc and the formation of the bar by placing these events on a common timeline and age scale. Although the retrograde GS/E merger has been shown to be important in producing the $\alpha$-bimodality, one caveat has not yet been addressed: whether a low-$\alpha$ disc formed after a retrograde GS/E merger could still have the long scale length observed for the Milky Way's low-$\alpha$ thin disc. Table~1 of \citet{Amarante2026} shows that the scale length of the disc formed in galaxies with a retrograde GS/E merger is significantly shorter than that observed in the Milky Way, while a prograde merger produces a scale length similar to that of the Milky Way but struggles to produce an $\alpha$-bimodality as strong as the Milky Way's. \citet{Zhang2026} also showed that an increase in the scale length of the star-forming disc after the GS/E merger is important in shaping the age--metallicity sequence of the low-$\alpha$ disc, similar to the sudden growth of the star-forming disc after the GS/E merger shown by \citet{Agertz2021_VI}. An important ingredient that might not be represented faithfully in all the models mentioned above is radial migration \citep{Sellwood_Binney2002, Minchev2013}, which has also been shown to help produce an $\alpha$-bimodality \citep[][]{Sharma2021, Chen2023, Zhang2025}. Joint modelling of how the GS/E merger and radial migration together shape the chemical evolution and the age--metallicity plane of the Milky Way is needed to establish the role of the GS/E in the emergence of the $\alpha$-bimodality and the low-$\alpha$ disc. The chemodynamical properties of the GS/E merger-induced starburst presented in this work could be a useful benchmark for these models.

\section{Discussion and conclusion}
\label{sec:conclusions}


We have characterised \Eos, a population of in-situ, low-$\alpha$ stars on halo-like orbits \citep{Myeong2022, Davies2025}. The chemical properties of Eos are similar to those of the most metal-poor stars in the low-$\alpha$ disc, but its dynamics are distinct from those of both the low-$\alpha$ disc and the GS/E debris (Figs~\ref{fig:3pops} and \ref{fig:vphi}). We traced its origin by comparing its chemodynamics with the star formation triggered by a GS/E-like merger in the Auriga halo Au18. Our main results are as follows.

\begin{enumerate}
    \item \new{The metal-poor branch of \Eos\ has a nitrogen dispersion comparable to that of the pre-disc Aurora and up to 1.7 times that of the low-$\alpha$ disc at the same metallicity. The Splash, a disc heated onto halo orbits and selected in the same way, shows no such excess over its parent disc, so the dispersion of \Eos\ was set at birth rather than by heating. \Eos\ therefore formed from poorly mixed gas, unlike the quiescent disc (Figs~\ref{fig:N_dispersion} and \ref{fig:bifurcation}).}

    \item \Eos\ is chemically and dynamically intermediate between the GS/E debris and the Milky Way's coeval in-situ stars, and the two properties track each other: stars with more GS/E-like chemistry (lower \alfe) are on more extended orbits (larger $r_{\rm apo}$, $z_{\rm max}$), and vice versa (Figs~\ref{fig:al_fe_orb} and \ref{fig:mg_al_plane}). This coupled gradient is consistent with a birth environment in which gas from the two galaxies was actively mixing.

    \item \Eos\ is older than the bulk of the low-$\alpha$ disc but younger than the Splash, a relative ordering that holds on both age scales; on the astroNN scale it is also younger than the high-$\alpha$ disc (Figs~\ref{fig:amr} and \ref{fig:bingo_age}). In the age--metallicity plane it extends the locally formed low-$\alpha$ sequence to older ages and lower metallicities, marking the onset of low-$\alpha$ disc formation.

    \item \Eos\ splits into a metal-poor branch (extended orbits, Aurora-like nitrogen dispersion) and a metal-rich branch (tighter, Splash-like orbits, disc-like dispersion) with distinct dynamics, pointing to a dual origin (Fig.~\ref{fig:bifurcation}).

    \item In Au18, the merger funnels the stripped progenitor gas into a lane of intermediate metallicity that rises towards the host. Stars born along it inherit intermediate abundances and the largest chemical dispersion of any merger-epoch population, and their birth sites map onto a present-day orbital gradient, reproducing both the chemical dispersion and the chemodynamical coupling seen in \Eos\ (Figs~\ref{fig:au18_lane_fehzmax} and \ref{fig:au18_Ndisp}).

    \item The lane starburst forms a halo-orbit population comparable in mass to the coeval Splash (Section~\ref{sec:sim_sfr}), so a chemically intermediate relic of this event is expected to survive in the local halo; restricted to the short window of incomplete mixing, its expected mass is similar to that estimated for Eos. 
\end{enumerate}

These results support a picture in which the metal-poor part of \Eos\ is the surviving, chemically intermediate relic of the GS/E merger-induced starburst, born mostly from the partially mixed gas along the lane bridging the GS/E and the Milky Way. Its intermediate chemistry and broad, Aurora-like nitrogen dispersion both follow from a single turbulent, poorly mixed birth episode, in contrast to the quiescent star formation that built the bulk of the low-$\alpha$ disc, and favour a halo-born origin along the gas lane over a pre-existing disc that was merely heated. This is consistent with the ``first sparks'' of clumpy star formation inferred by \citet{An2025}, the GS/E-associated origin proposed by \citet{Myeong2022} and \citet{Davies2025}, and the merger-induced starburst timed by \citet{Ciuca2024} and \citet{Laporte2026}. The metal-rich (roughly $[\rm Fe/H]>-0.6$) part of the non-rotating low-$\alpha$ stars, in contrast, shares the disc-like nitrogen dispersion of the low-$\alpha$ disc (Fig.~\ref{fig:bifurcation}); its chemistry alone therefore cannot discriminate between formation in the merger-induced starburst and an origin in the pre-existing low-$\alpha$ disc, dynamically heated onto halo orbits, as proposed by \citet{Borbolato2026}.

Sitting at the oldest, most metal-poor end of the low-$\alpha$ sequence and following the chemical evolution trends of the low-$\alpha$ stars, \Eos\ plausibly marks the onset of low-$\alpha$ disc formation and preserves a chemodynamical record of the in-situ response to the GS/E merger. We stress that it most likely traces, rather than builds, the low-$\alpha$ disc, since a single GS/E-like merger cannot chemically sustain the whole sequence \citep[][]{Orkney2026a_bimodality}. \Eos\ nonetheless remains a largely unexploited probe of the merger epoch. A homogeneous, carefully calibrated age for \Eos\ would turn it into a clock for the last pericentric passage of the GS/E, complementary to the Taina globular clusters \citep{Laporte2026}, while the larger and cleaner samples expected from SDSS-V, WEAVE and 4MOST will pin down its bifurcation, its chemical dispersion, and the chemodynamical gradients that encode the gas mixing during the merger. Together, these measurements would make \Eos\ a quantitative benchmark for the star formation triggered by the Milky Way's last significant merger.

\section*{Acknowledgements}
HZ thanks the Science and Technology Facilities Council (STFC) for a PhD studentship (grant number 2888170). VB acknowledges support from the Leverhulme Research Project Grant RPG-2021-205: ``The Faint Universe Made Visible with Machine Learning''. JASA is supported by the National Natural Science Foundation of China under grant Nos. 12233001, 12533004, by the National Key R\&D Program of China under grant No. 2024YFA1611602, by a Shanghai Natural Science Research Grant (24ZR1491200), by the ``111'' project of the Ministry of Education under grant No. B20019, by the China Manned Space Program with grant Nos. CMS-CSST-2025-A08, CMS-CSST-2025-A09 and CMS-CSST-2025-A11, and in part by Office of Science and Technology, Shanghai Municipal Government (grant Nos. 24DX1400100, ZJ2023-ZD-001). SGK thanks the Cambridge Trust and Trinity College, Cambridge for PhD funding.

\section*{Data availability}
All data used in this work are publicly available. Our selected sample of Eos stars is available at: https://zenodo.org/records/24346312
\bibliographystyle{mnras}
\bibliography{bibliography}

\appendix

\section{Age distribution of Eos with other spectroscopic ages}
\label{appendix:bingo_age}
\begin{figure}
    \centering
    \includegraphics[width=0.99\linewidth]{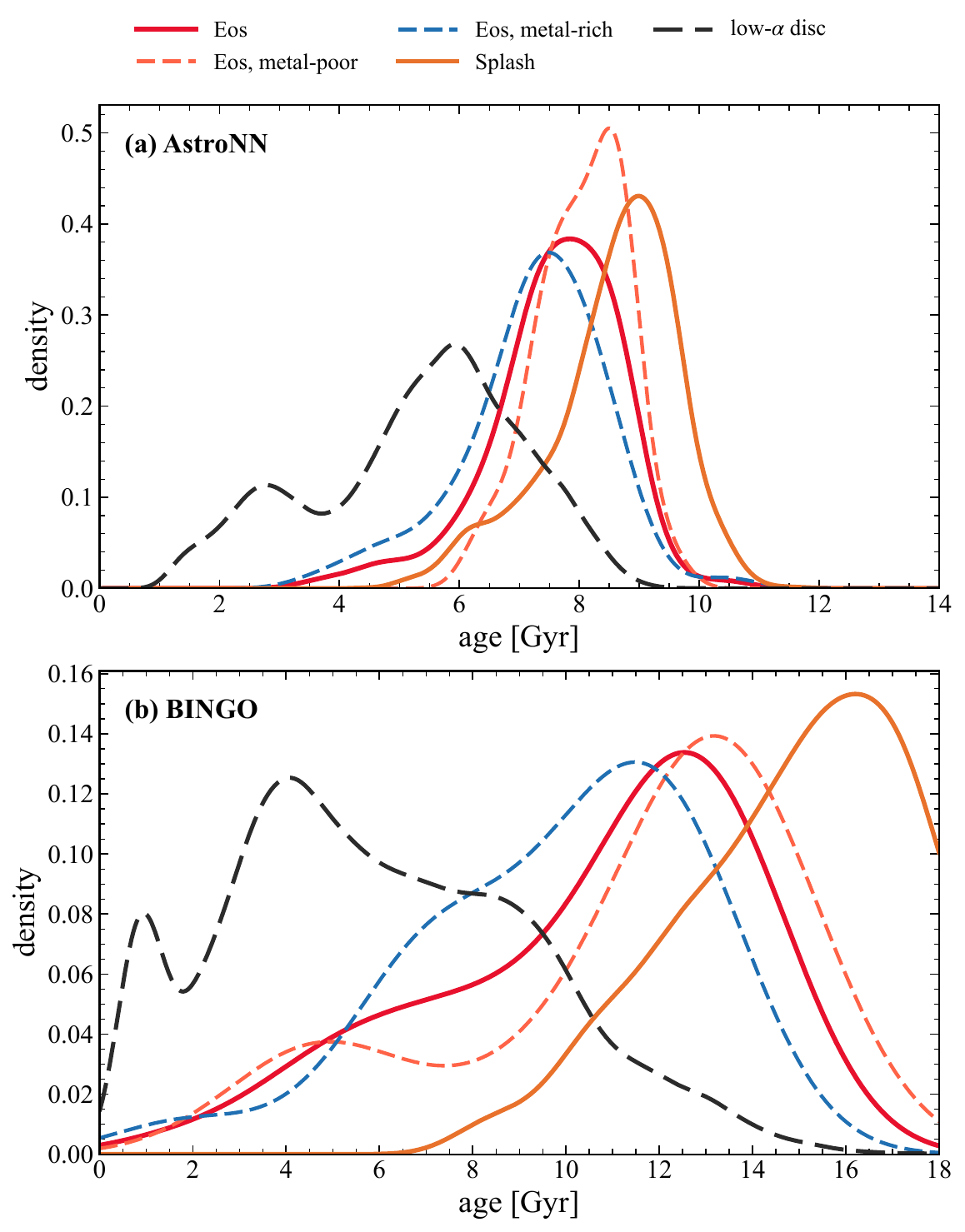}
    \caption{Age distributions of Eos and the Milky Way populations on two spectroscopic age scales: \textit{(a)} astroNN and \textit{(b)} BINGO \citep{Ciuca2021}. Shown are Eos (red), its metal-poor (tomato, dashed) and metal-rich (blue, dashed) branches, the Splash (orange) and the low-$\alpha$ disc (dark, dashed). The relative ordering (Eos older than the low-$\alpha$ disc but younger than the Splash) is preserved on both scales, although the absolute ages differ by several Gyr, so the ages should be read only in a relative sense.}
    \label{fig:bingo_age}
\end{figure}

In Section~\ref{sec:eos_age} we established the relative age ordering of \Eos\ using the astroNN ages \citep{Leung2019}. Because spectroscopic ages carry catalogue-dependent systematics, particularly for old stars, we repeat the measurement here with a second, independent age scale, the \texttt{BINGO} catalogue \citep{Ciuca2021}. Figure~\ref{fig:bingo_age} shows the age distributions of \Eos, its metal-poor and metal-rich branches (Section~\ref{sec:eos_bifurcation}), the Splash and the low-$\alpha$ disc on the two scales, applying a $\sigma_{\rm age}/{\rm age}<0.15$ quality cut to the astroNN panel and the \citet{Ciuca2024} quality cuts to the raw (uncalibrated) BINGO ages (122 and 46 Eos stars pass these cuts, respectively). The relative ordering (\Eos\ older than the low-$\alpha$ disc but younger than the Splash) is preserved in both catalogues, showing that this ordering, which underpins our interpretation of \Eos\ as the onset of the low-$\alpha$ disc, is not an artefact of a single age scale. The absolute ages nevertheless differ by several Gyr (the mode of \Eos\ shifts from $\approx8$ to $\approx12$~Gyr between astroNN and BINGO), which is why we use the ages only in a relative sense throughout. In particular, the raw BINGO ages of the oldest populations are unreliable in an absolute sense (the Splash peaks near $16$~Gyr), reinforcing that only the relative ordering in Figure~\ref{fig:bingo_age} should be read. We refer the reader to Figure~8 of \citet{funakoshi2026} for the [Fe/H]--[Mg/Fe] distribution of the Milky Way populations coeval with Eos on the BINGO age scale: stars with ages equal to the age mode of Eos lie mainly on the high-$\alpha$ thick disc, consistent with our result in Figure~\ref{fig:mg_al_plane} based on astroNN ages. Further work that calibrates different age-measurement methods onto a common scale, along the lines of \citet{funakoshi2026}, would allow a better timeline of the Milky Way's evolution to be drawn.

\section{Nitrogen abundance dispersion during the GS/E-like merger}
\label{appendix:Au18_sigmaN}

\begin{figure*}
    \centering
    \includegraphics[width=0.99\linewidth]{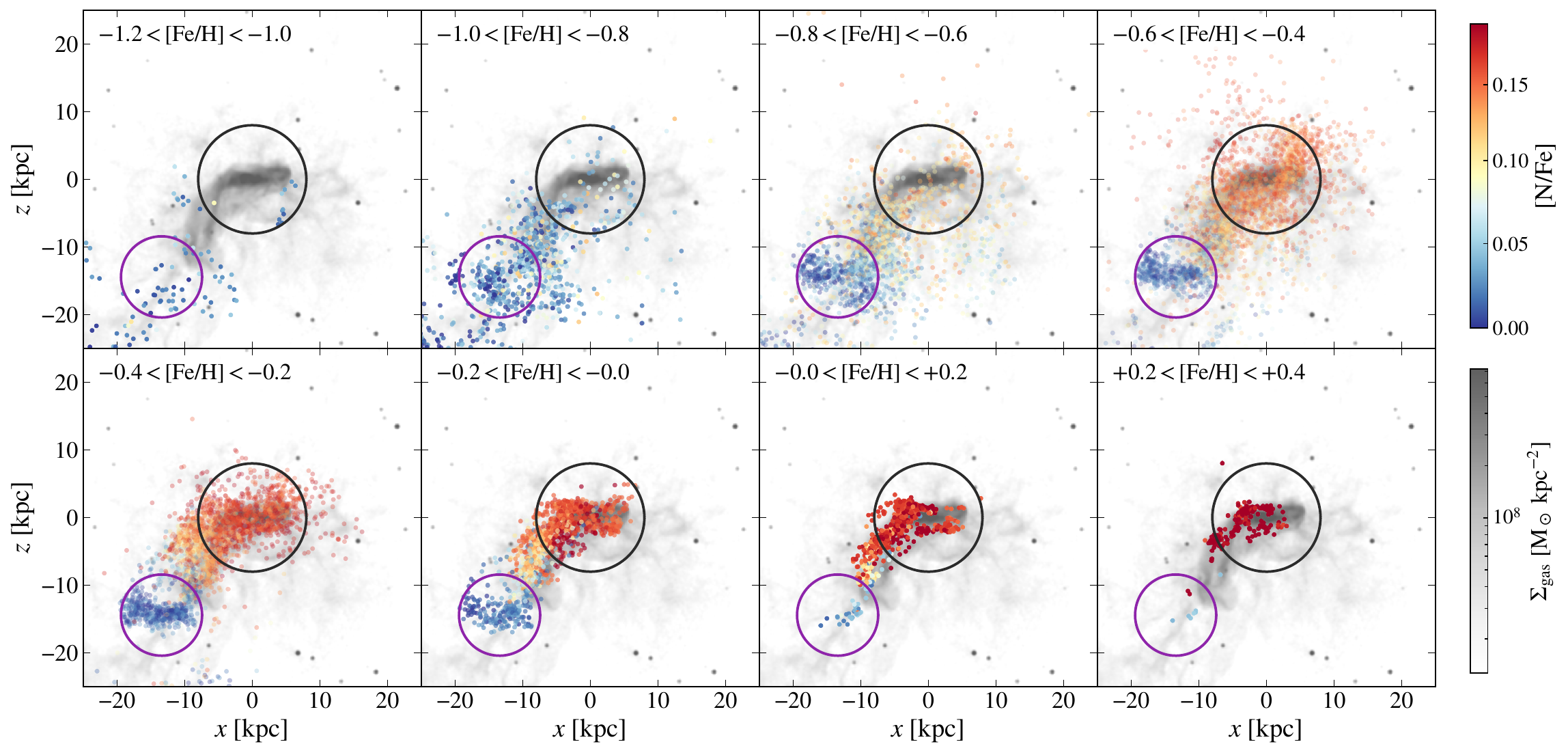}
    \caption{Edge-on maps of the halo-born stars in Au18 in bins of metallicity (labelled in each panel), coloured by [N/Fe], over the gas surface density (grey). The circles mark the GS/E analogue (purple) and host (dark) apertures. The large scatter in [N/Fe] at a given position, strongest along the lane between the two galaxies and at intermediate metallicity, illustrates the poorly mixed gas from which the lane stars form.}
    \label{fig:Au18_sigmaN}
\end{figure*}

Section~\ref{sec:au18_lane} quantified the nitrogen dispersion of the stars born during the GS/E-like merger in Au18 (Figure~\ref{fig:au18_Ndisp}) and showed that the stars forming along the gas lane carry the largest scatter. The middle panel of Figure~\ref{fig:au18_Ndisp} shows a low-high-low trend of nitrogen dispersion with metallicity for the halo-born stars. Figure~\ref{fig:Au18_sigmaN} visualises the spatial origin of this dispersion and of the trend. We take the same halo-born sample and plot it edge-on in bins of [Fe/H], colouring each star by its [N/Fe] and overlaying it on the gas surface-density map. At a fixed position, and most strongly along the lane connecting the two galaxies and at intermediate metallicity, [N/Fe] varies substantially from star to star, directly illustrating the poorly mixed gas out of which the lane stars form. The low-high-low trend in $\sigma_{\rm [N/Fe]}$ vs [Fe/H] arises from the correlation between the position of a star and its metallicity. The most metal-poor and most metal-rich stars born on halo orbits lie close to the GS/E progenitor and the main host, respectively. Because the nitrogen dispersion is high only for stars formed along the lane, at intermediate metallicity, this coupling between spatial location and metallicity translates into a low-high-low trend in nitrogen dispersion as a function of metallicity. This spatially resolved view complements the binned dispersion of Fig.~\ref{fig:au18_Ndisp} and provides the simulation counterpart to the large nitrogen spread observed in \Eos\ (Section~\ref{sec:eos_nitrogen}).

\section{Dual origin of Eos}
\label{appendix:dual_origin}

\begin{figure*}
    \centering
    \includegraphics[width=\linewidth]{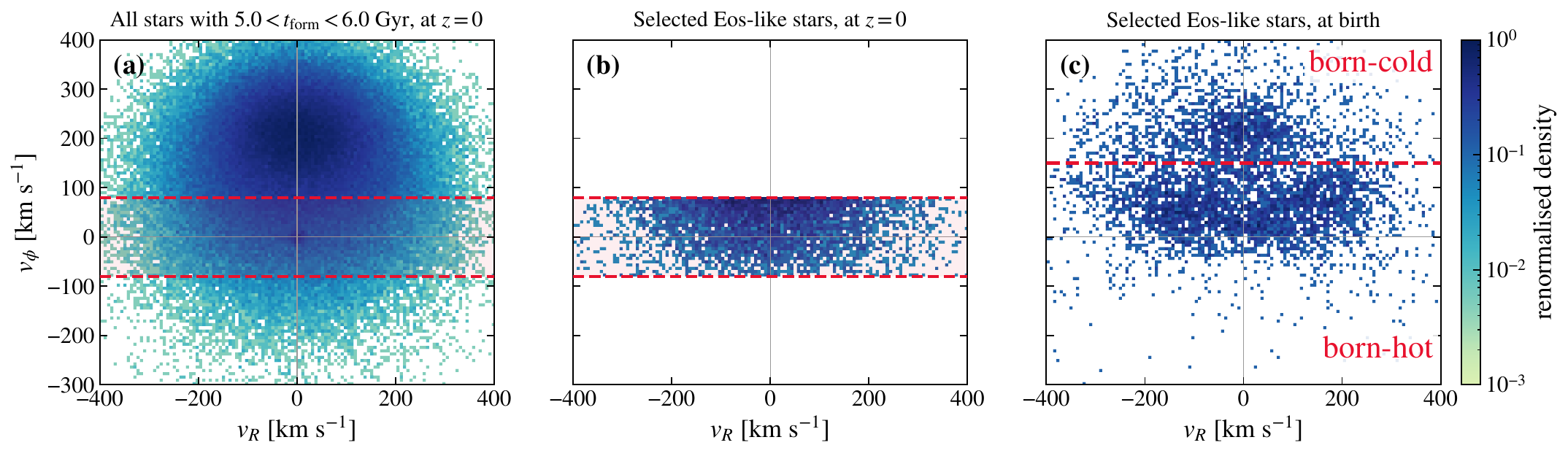}
    \caption{Birth and present-day kinematics of the in-situ stars formed within 1~Gyr after the last pericentric passage of the GS/E-like merger in Au18 ($5<t_{\rm form}<6$~Gyr). \textit{(a)} Present-day $v_R$--$v_\phi$ distribution of all such stars, dominated by the rotating disc at $v_\phi\approx200~\kms$. \textit{(b)} Present-day kinematics of the Eos-like subset, selected to have halo orbits today ($|v_\phi|<80~\kms$, $e>0.7$, $z_{\rm max}>2$~kpc). \textit{(c)} Birth kinematics of this subset, which separate into a born-cold population (low $v_R$, high $v_\phi$) and a born-hot population (large $v_R$ dispersion, intermediate $v_\phi$), split at $v_\phi=150~\kms$ and contributing 30 and 70 per cent, respectively. The red dashed lines mark $|v_\phi|=80~\kms$ in (a) and (b) and $v_\phi=150~\kms$ in (c).}
    \label{fig:au18_vr_vphi}
\end{figure*}

\begin{figure}
    \centering
    \includegraphics[width=\linewidth, height=1.3\linewidth]{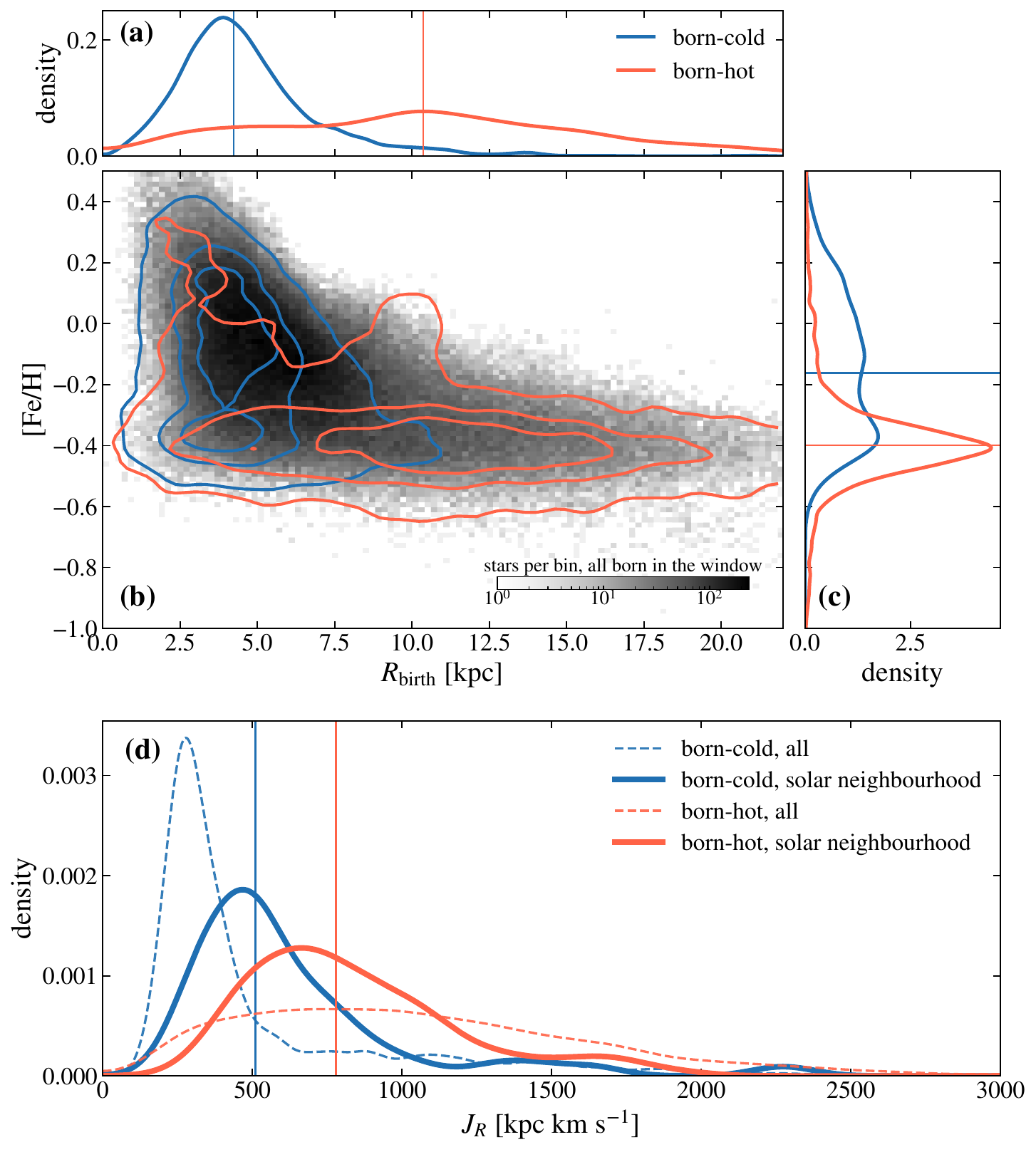}
    \caption{Birth radius, metallicity and radial action of the born-cold (blue) and born-hot (red) Eos-like populations in Au18. \textit{Centre:} Birth cylindrical radius versus [Fe/H]; the grey density shows all stars born in the merger window and the contours (enclosing 30, 60 and 90 per cent) show the two populations, with their marginal distributions in the \textit{top} and \textit{right} panels. \textit{Bottom:} The radial-action $J_R$ distributions of the two populations for the whole galaxy (dashed) and after a solar-neighbourhood selection (solid). The born-hot population forms at larger radii, is more metal-poor, and has systematically higher $J_R$, mirroring the metal-poor Eos branch. Thin lines mark the medians (in the bottom panel, of the solar-neighbourhood selections).}
    \label{fig:au18_jr}
\end{figure}

In the main text, we connected the Eos population to the merger-induced halo starburst. However, this does not address the bifurcation of the Eos population in the [Fe/H]--[Mg/Fe] plane discussed in Section~\ref{sec:eos_bifurcation}, where the two branches also have distinct kinematic properties and therefore likely have different origins. In this section, we discuss a potential origin of this bifurcation using its simulated counterpart.

Figure~\ref{fig:au18_vr_vphi} investigates the birth kinematics of the in-situ stars born around the time of the merger that have halo-like kinematics in the present-day ($z=0$) snapshot. The left panel shows the logarithmic density in the present-day $v_R-v_\phi$ plane of all the stars formed within 1~Gyr after the last pericentric passage. The majority of stars lie around $v_R\approx0$\,\kms\ and $v_\phi \approx 200$\,\kms; these are the classical disc stars. We select those with halo-like kinematics today, with $|v_\phi|<80$\,\kms, $e>0.7$ and $z_{\rm max}>2~\rm kpc$, similar to the selection we applied to the APOGEE sample. The last cut, on $z_{\rm max}$, helps to remove stars on bar orbits, which have low rotation speeds owing to the bar streaming motion but always stay close to the plane. The kinematics of the selected Eos-like population in the present-day Au18 snapshot are shown in the middle panel. In total, these stars make up $\approx3$ per cent of all the stars born in this time window. Note that this selection aims at a clean, rather than complete, sample of halo-like candidates. The right panel shows the birth kinematics of the selected Eos-like population, traced by the snapshot closest to each star's birth time. Two main clouds are visible in this panel: one with small $v_R$ and high $v_\phi$ (a born-cold population) and one with intermediate $v_\phi$ and a large $v_R$ dispersion (a born-hot population). These correspond, respectively, to stars born in the disc and later splashed by the GS/E merger, and to stars born in the halo. We split the two at $v_\phi=150$\,\kms. Under these cuts, the born-cold population contributes 30 per cent of the Eos-like stars, and the born-hot population contributes the remaining 70 per cent. The born-hot stars have intermediate rotation at birth but $|v_\phi|<80~\kms$ today; this is likely because their birth $v_\phi$ is measured in the disc frame at formation, while the disc spin axis swings by $\approx50^\circ$ across the merger coalescence, so that rotation defined in the final frame is largely washed out.

We further compare the observables of these two populations born with different kinematics. Figure~\ref{fig:au18_jr} compares the metallicity and radial action distributions of the born-cold (blue) and born-hot (red) stars. The centre panel shows metallicity versus birth cylindrical radius. The born-cold population has a higher metallicity than the born-hot stars (their medians differ by $0.2~\rm dex$). This is because the born-cold stars have birth radii more than a factor of two smaller than those of the born-hot population, and the inner galaxy is more metal-rich, as seen in the grey background distribution of all the stars born during this time span. The lower panel shows the radial action ($J_R$) distributions. The dashed lines show the $J_R$ distributions of the born-cold and born-hot populations for the whole galaxy. The born-cold population is strongly clustered at small $J_R$, while the born-hot population has a much wider $J_R$ distribution that extends to high values. However, the radial action distribution is typically sensitive to the selection function. We therefore apply a spatial selection that removes stars more than 4~kpc from a fiducial solar position, placed 8~kpc from the galactic centre at a randomly chosen azimuth. This roughly mimics the spatial distribution of our observational sample. The $J_R$ distributions after this spatial selection are shown as solid lines. The $J_R$ distribution of the born-cold population is no longer concentrated at small values and now has a median of $\approx500~\rm kpc~km~s^{-1}$. The $J_R$ distribution of the born-hot population shifts towards smaller values, with the median now at $\approx750~\rm kpc~km~s^{-1}$. 

Comparing the chemodynamical properties of the born-cold and born-hot populations in Au18 with the observed Eos population and its bifurcation, we find strong similarities. The metal-poor branch of the Eos population in Figure~\ref{fig:bifurcation} also shows larger $J_R$ and $r_{\rm apo}$ than the metal-rich branch. The similarity of their chemodynamics suggests a possible origin of the Eos bifurcation: the metal-poor branch of Eos consists of stars born along the gas lane between the GS/E progenitor and the Milky Way, and the metal-rich branch consists of low-$\alpha$ disc stars that formed early, during the GS/E merger, and were splashed onto halo orbits. A lane origin could explain the large apocentres and high nitrogen dispersion of the metal-poor Eos stars. In this picture, both the gas mixing and the rapid starburst raise the [Mg/Fe] and [Al/Fe] abundances above those of the GS/E stars. The metal-rich Eos population indeed has a nitrogen dispersion consistent with that of the low-$\alpha$ disc stars with high azimuthal velocity at the same metallicity, and apocentres consistent with those of the high-$\alpha$ Splash stars.

\bsp	
\label{lastpage}
\end{document}